\documentclass[useAMS,usenatbib,times,letter,amssymb]{mn2e}
\usepackage{graphicx, float, amsmath,epsfig,times,amssymb,color,verbatim}
\def\be{\begin{equation}}
\def\ee{\end{equation}}
\def\bea{\begin{eqnarray}}
\def\eea{\end{eqnarray}}
\def\ATrue{0.7}
\newcommand{\bm}[1]{\mbox{\boldmath{$#1$}}}   

\title[Clustering with flux magnification and shear]{On the complementarity of galaxy clustering with cosmic shear and flux magnification}

\author[C. Duncan et al.]{Christopher A. J. Duncan$^{1}$\thanks{Email: cajd@roe.ac.uk}, Benjamin Joachimi$^{1}$, Alan F. Heavens$^{2}$, Catherine Heymans$^{1}$,\newauthor Hendrik Hildebrandt$^{3}$\\
$^{1}$Scottish Universities Physics Alliance, Institute for Astronomy, University of Edinburgh, Royal Observatory, Blackford Hill, Edinburgh, EH9 3HJ, UK\\
$^{2}$Imperial Centre for Inference and Cosmology, Imperial College London, Blackett Laboratory, Prince Consort Road, London, SW7 2AZ, UK.\\
$^3$Argelander Institute for Astronomy, University of Bonn, Auf dem Hugel 71, 53121 Bonn, Germany}

\par

\begin{document}

\date{Accepted}

\pagerange{\pageref{firstpage}--\pageref{lastpage}} \pubyear{2013}

\maketitle

\label{firstpage}

\begin{abstract}
With the wealth of forthcoming data from wide--field surveys, it is more important than ever to understand the full range of independent probes of cosmology at our disposal.  Here we explore the potential for galaxy clustering and cosmic shear, separately and in combination, including the effects of lensing magnification. We show that inferred cosmological parameters may be biased when flux magnification is neglected. Results are presented for Stage III ground--based and Stage IV space--based photometric surveys, using slopes of the luminosity function inferred from the CFHTLenS catalogue. We find that combining with clustering improves the shear Dark Energy Task Force--like figure of merit by a factor of 1.33 using only auto--correlations in redshift for the clustering analysis, rising to 1.52 when cross--correlations are also included. The further addition of galaxy--galaxy lensing gives increases in the shear figure of merit by a factor of 2.82 and 3.7 for each type of clustering analysis respectively. The presence of flux magnification in a clustering analysis does not significantly affect the precision of cosmological constraints when combined with cosmic shear and galaxy--galaxy lensing. However if magnification is neglected, inferred cosmological parameter values are biased, with biases in some cosmological parameters larger than statistical errors.
\end{abstract}


\section{Introduction}

As light from distant galaxies propagates through the Universe, its path can be deflected by the local matter distribution, an effect called gravitational lensing. As a result, when we view distant galaxies we observe both a change in the shape of its image, as well as a change in its position and size. Further, as gravitational lensing conserves surface brightness, this change in size also causes a change in the observed flux of a lensed source. Cosmic shear is the study of the observed change in shape, and statistical analysis using galaxy ellipticities has proven a very promising tool for probing cosmology \cite[reviewed in ][]{Bartelmann:2001p431, Munshi:2006p1131}. However, as shear analysis requires accurate shape information, its measurement, sensitive to the Point Spread Function (PSF), pixelisation \citep{Bridle:2009p1030, Kitching:2012p2018} and noise bias \citep{Bernstein:2002p2069, Hirata:2004p2070, Melchior2012,Refregier2012}, has proven a particularly difficult task .

With many current and forthcoming large scale surveys such as CFHTLenS\footnote{http://http://www.cfhtlens.org}, DES\footnote{http://www.darkenergysurvey.org}, KiDS\footnote{http://www.astro-wise.org/projects/KIDS}, HSC\footnote{http://www.naoj.org/Projects/HSC/index.html} and Euclid\footnote{http://www.euclid-ec.org}\citep{Laureijs:2011p2021}, it is becoming more important than ever to understand what gains there are to be made in exploiting the other half of the lensing signal. 

Direct observations of the induced change in size of a lensed body is the most obvious measure of the magnification effect, and magnification has been successfully detected using a combination of the change in sizes of a lensed population of galaxies and a change in their magnitudes in \cite{Schmidt:2012p1106}. However, due to the nature of the measurements required to detect size change due to magnification, it suffers from many of the same systematics listed for cosmic shear above. In \cite{Casaponsa:2013p1480} it was shown that these systematics may limit the use of size change for ground--based surveys where the effects of PSF are largest. However for space--based surveys with smaller PSF and higher signal--to--noise ratio, the statistical power of size magnification can rival cosmic shear, and may be an excellent complement to shear analyses \citep{Heavens:2013p1550}.

Another recent technique to use the magnification part of the lensing signal uses the fundamental plane to relate the effective radius of a galaxy, which is altered by magnification, to the galaxy's surface brightness and stellar velocity dispersion, both of which remain unaltered \citep{Huff:2011p1392}.

The most common technique called flux magnification (or magnification bias) uses fluctuations in the observed number density of sources as a probe of cosmic magnification. Observationally, the observed number density of sources is altered due to magnification in two ways, the dilution of sources as the solid angle behind the lens is stretched, and the (de--)amplification of sources as their fluxes are (de--)magnified (below) above the survey flux limit. Lensing can therefore induce non--vanishing number density contrast correlations between a background distribution of sources and foreground large scale structure, which are sensitive to cosmology through the distribution of matter and its evolution, and distance measures.

The early history of using flux magnification to measure magnification proved controversial with early attempts at measuring changes in number density of a background set of quasars due to foreground large scale structure commonly in disagreement, and with measurements that gave amplitudes of correlations far larger than theoretical predictions (\citealt{Scranton:2005p1124} provides a concise summary of early literature).
 However, the successful measurement of number density contrast correlations between background quasars and foreground galaxies using the Sloan Digital Sky Survey \citep{Scranton:2005p1124} and the 2dF \citep{Myers2dF2005}, and later using high redshift Lyman break galaxies as sources in CARS \citep{Hildebrandt:2009p845}, laid the basis for the use of magnification as a probe of cosmology and large scale structure.

In contrast to cosmic shear analysis, which has seen a recent concentrated effort by the lensing community to remove or understand measurement or statistical systematics, flux magnification as a probe of cosmology is relatively less mature. The reasons for this are simple, for a given sample of galaxies the signal--to--noise ratio for flux magnification is expected to be smaller, as the shot noise in the shear case is reduced by factor of the square of the intrinsic ellipticity dispersion for each ellipticity component. However it should be noted that particularly in the case of ground--based surveys we expect to be able to use a greater number of galaxies in a magnification analysis (provided accurate photometry is determined for these galaxies) than for cosmic shear, as the measurement itself is easier and does not require accurate shape information. This should go some way to offsetting the discrepancy in signal--to--noise ratio between shear and flux magnification. 

The use of flux magnification as a probe of cosmology is mainly limited by errors in determining the photometric redshift of sources. In particular, the amplitude of the number density contrast correlation from magnification is smaller than that induced by the intrinsic clustering of galaxies due to their dark matter environment, making it difficult to disentangle these signals in the presence of photometric scatter, causing the intrinsic clustering of spatially close populations to be mis--interpreted as a magnification signal, and giving spurious results. Previous analyses have attempted to remove most of this contamination by choosing carefully selected foreground and background populations which are spatially disjoint \cite[such as][]{Hildebrandt:2009p845, VanWaerbeke:2010p8}, or using the nulling technique \citep{Heavens:2011p629}. Further, dust extinction and fluctuations of the magnitude zero points over the survey area can produce fluctuations in number density that mimic the magnification signal, and remain relatively unexplored.

In this paper we consider the use of induced correlations in number density between tomographically binned samples of galaxies as a probe of cosmology using a Fisher matrix analysis. We critically focus on what potential gains are to be made when combining a shear and magnification analysis, and then turn to what biases to inferred cosmological parameters would be introduced if magnification was incorrectly neglected in a clustering analysis. The theoretical predictions for the number density power spectra are set out in Section \ref{Sec:2ptCorrelations}.  In Section \ref{Sec:Modelling} we detail the modelling of galaxy bias and galaxy distributions with photometric redshift errors, and information on number counts taken from public CFHTLenS catalogues \citep{Heymans:2012p1653,Erben:2012p1556} are used to motivate physical ranges for the slope of the luminosity function. Results are presented in Section \ref{Sec:Results}, detailing forecasts using a Fisher matrix formalism for two types of current and future survey, and work investigating how inferred cosmological parameters are biased in a clustering analysis if the flux magnification contribution to number density fluctuations is incorrectly assumed to be zero. Concluding remarks are presented in Section \ref{Sec:Conclusions}.
	
\section[]{Theory}\label{Sec:2ptCorrelations}
\subsection{Number density and cosmic shear statistics}

The action of gravitational lensing by a foreground matter distribution causes two effects on the observed images of a background source galaxy: A change in the observed shape of a galaxy, commonly parameterised in a change in the measured ellipticity ($\epsilon$) of said galaxy caused by gravitational shear, and a change in the size, flux and consequently the number density of sources caused by gravitational magnification. In this paper we consider the use of both galaxy ellipticity measurements, galaxy clustering and the change in observed number density of sources in a flux limited survey as probes of cosmology. 

We consider the observed projected ellipticity of sources using the estimator
\be\label{Eqn:EllipticityEstimator}
\epsilon^{(i)}({\bm\theta}) = \gamma^{(i)}({\bm\theta}) + \epsilon^{(i)}_{\rm rn}(\bm{\theta})
\ee
where ${\bm \theta}$ denotes the two--dimensional position on the sky, the gravitational shear of the source is denoted by $\gamma$, and $\epsilon_{\rm rn}$ is a random stochastic element that contributes to the noise. Superscripts in brackets label the redshift bin of tomographically binned sources. The convergence $\kappa$,  is defined as 
\be\label{eqn:ProjectedConvergence}
\kappa_{y}(\bm{\theta}) = \int_0^{\chi_H} d\chi\; q_{y}^{(i)}(\chi) \delta(\bm{\theta},\chi) 
\ee
with the dark matter over--density denoted by $\delta$, and the weight given as \citep{Bartelmann:2001p431}
\be\label{eqn:ConvergenceWeight}
q_{y}^{(i)}(\chi) = \frac{3H_0^2\Omega_m}{2c^2}\frac{f_K(\chi)}{a(\chi)}\int^{\chi_H}_\chi d\chi'\; p_{y}^{(i)}(\chi')\frac{f_K(\chi'-\chi)}{f_k(\chi')}.
\ee
where $a$ is the dimensionless scale factor, and $\chi_{H}$ the co--moving particle horizon. The convergence has identical two--point statistics in Fourier space to the gravitational shear $\gamma$, and we therefore use it as the shear observable. Here we have used subscript $y = \{ {\rm s,M}\}$ to differentiate between convergence measured from shape and photometry samples. The galaxy comoving distance probability distribution for tomographic redshift bin $i$ is denoted by $p^{(i)}(\chi)$ normalised such that $\int d\chi \;p^{(i)}(\chi)  = 1$ for all redshift bins. 

Deflection of light by intervening matter causes the observed number density of sources to be changed in two ways:
\begin{enumerate}
\item{The solid angle behind the lens is increased by a factor of $\mu$ (where $\mu$ is the local magnification factor), thus the observed position of sources is changed leading to a dilution of sources behind a foreground over--density}
\item{The observed size of the source is changed, leading to a change in the observed flux of the source as surface brightness is conserved by lensing. The observed number density of sources may then change in a flux--limited survey, as sources are (de--)amplified across this flux limit $f$. This is equivalent to a local effective change in the flux limit of the survey, which changes to $f/\mu$.}
\end{enumerate}

These two effects then modify the observed number density of sources at position ${\bm \theta}, \chi$ as 

\begin{equation}\label{eqn:Lensed-UnlensedNumberCounts}
n(>f,{\bm \theta}, \chi) = \frac{n_0(>f/\mu({\bm\theta}, \chi),{\bm\theta})}{\mu({\bm\theta})}.
\end{equation}

Approximating the unlensed number counts as following a power law at the faint end, $n_0(>f) \propto f^\alpha$, the observed number counts are given by
\bea\label{eqn:LensedNumberCounts_Kappa}
n(>f, {\bm \theta}, \chi) &=& \mu^{\alpha(f)-1}n_0(>f,{\bm \theta}, \chi)\nonumber\\ &\approx& \{1+2[\alpha(f)-1]\kappa_{\rm M}(\bm{\theta})\}n_0(>f,{\bm \theta}, \chi), 
\eea
where the weak lensing approximation $\mu \approx 1+2\kappa$ has been used and the result Taylor--expanded around $\kappa = 0$. 

From Equation (\ref{eqn:LensedNumberCounts_Kappa}) it is clear that when $\alpha=1$ the overall magnification effect does not cause a change in the observed number density of sources, as the dilution of sources is perfectly balanced by the increased number of galaxies caused by the amplification of sources over the flux limit of the survey. Alternatively, when $\alpha \ne 1$ there will be an overall increase/reduction in the observed number of sources. In terms of magnitudes,
\be\label{Eqn:Alpha_Magnitude}
\alpha(i_{\rm AB}) = 2.5\frac{{\rm d}\:{\log_{10}\:n(>i_{\rm AB})}}{{\rm d}i_{\rm AB}}.
\ee
where we have quoted an i--band, AB magnitude $i_{\rm AB}$, chosen here as we will use this passband when analysing CFHTLenS data in Section \ref{Sec:CFHT}. Defining the number density contrast as $\delta n = (n-n_0)/n_0$, Equation (\ref{eqn:LensedNumberCounts_Kappa}) gives the fluctuation in observed number density due to magnification:
\be\label{eqn:NDensityContrastFluctuationMag}
\delta n_m(\bm{\theta}) = 2(\alpha-1)\kappa_{M}(\bm{\theta}).
\ee 
 The observed number density contrast is then
\be\label{eqn:NumberDensityContributions}
\delta n^{(i)}(\bm{\theta}) = \delta n_m^{(i)}(\bm{\theta}) + \delta n_g^{(i)}(\bm{\theta}) + \delta n_{rn}^{(i)}(\bm{\theta}),
\ee
with being $\delta n_g$ the contribution from the intrinsic clustering of the sources, and $\delta n_{rn}$ a random stochastic element. 

The projected number density contrast due to intrinsic clustering is related to the three--dimensional number density fluctuations $\delta_g$ by
\be\label{eqn:Projected_NumberDensity}
\delta n_g(\bm{\theta}) = \int_0^{\chi_H} d\chi\; p_{M}^{(i)}(\chi) \delta_g(\bm{\theta},\chi).
\ee

As number density fluctuations and source ellipticity vanishes when averaged over large scales, we consider two--point correlations of these quantities. In particular, we consider the two--point correlation of the Fourier coefficients of the convergence and number density contrast, related to the power spectrum $P$ by:
\be\label{eqn:PSDef}
\langle x^{(i)}(\bm{\ell})y^{(j)}(\bm{\ell'})\rangle = (2\pi)^2\delta_D(\bm{\ell}-\bm{\ell'})P^{(ij)}_{xy}(\bm{\ell}),
\ee
for variables $x,y = \{\delta n, \kappa_s\}$. The two--dimensional Dirac delta function $\delta_D(\bm{\ell}-\bm{\ell'})$ illustrates the non--mixing of angular wavenumber ($\ell$) modes due to homogeneity on the sky, and we make a flat sky approximation.  We construct three observables:
Firstly, the `galaxy clustering' power spectra, constructed from position--position correlations including flux magnification contributions:
\bea
P^{(ij)}_{\delta n\delta n}(\bm{\ell}) &=& P^{(ij)}_{mm}(\bm{\ell}) +P^{(ij)}_{gg}(\bm{\ell}) + P^{(ij)}_{mg}(\bm{\ell}) + P^{(ij)}_{gm}(\bm{\ell})   \nonumber\\
 & & + \delta_K^{ij}P^{SN}_{\delta n}. \label{eqn:NumberDensityContrastPS}
 \eea
 Secondly, `cosmic shear' power using ellipticity--ellipticity correlations:
 \be
P^{(ij)}_{\epsilon\epsilon}(\bm{\ell})  = P^{(ij)}_{\kappa_S\kappa_S}(\bm{\ell}) + \delta_K^{ij}P^{SN}_{\epsilon}. \label{eqn:Galaxy-Galaxy_PS} \\
\ee
Thirdly, `galaxy--galaxy lensing' power spectra, using position--ellipticity correlations:
\be
P^{(ij)}_{\epsilon\delta n}(\bm{\ell}) = P^{(ij)}_{\kappa_S g}(\bm{\ell}) + P^{(ij)}_{\kappa_S m}(\bm{\ell}),\label{eqn:NumberDensityContrastPS-GGLensing}
\ee
where $\delta_K^{ij}$ is the Kronecker symbol. For notational convenience, we have altered subscripts so that subscript `$m$' denotes the fluctuation in number density due to flux magnification (formally $\delta n_m$) and subscript `$g$' the fluctuation due to intrinsic clustering (formally $\delta n_g$). The stochastic term for the number density contrast and shear are uncorrelated with the other quantities and only contribute to the shot noise ($P^{SN}$) in the autocorrelation term. Readers should note that the presence of flux magnification modifies not only the clustering power spectra, but also adds an additional term to the galaxy--galaxy lensing power spectra.

The fluctuation due to intrinsic clustering is related to the matter over--density via a bias term ($b$) that can be scale-- or distance-- dependent so that the intrinsic clustering contribution to the power spectrum is given by
\bea
P_{\delta_g\delta_g}(\bm{k},z) &=& b^2(\bm{k},z)P_{\delta\delta}(\bm{k},z),\\
P_{\delta_g\delta}(\bm{k},z) \;&=& b(\bm{k},z)r(\bm{k},z)P_{\delta\delta}(\bm{k},z),
\eea
where $r(k,z)$ is a stochastic bias which we take to be unity for the remainder of this paper. Here, $P_{\delta \delta}$ denotes the three--dimensional matter density power spectrum, and $P_{\delta_g\delta_g}$ the three--dimensional intrinsic clustering number density contrast power spectrum. In this work, the matter power spectrum is modelled using \cite{Eisenstein:1998p1135} transfer functions with \cite{Smith:2003p843} non--linear corrections.

All power spectra terms for projected quantities are related to the three--dimensional dark matter power spectra using the Limber approximation in the flat sky limit. The contributions to the number density contrast power spectra in Equation (\ref{eqn:NumberDensityContrastPS}) are then
\bea
P^{(ij)}_{mm}(\bm{\ell}) &=& 4(\alpha^{(i)}-1)(\alpha^{(j)}-1)P^{(ij)}_{\kappa_M\kappa_M}(\bm{\ell}),\\
P^{(ij)}_{\kappa_{\rm s} m}(\bm{\ell}) &=& 2(\alpha^{(j)} - 1)P_{\kappa_{\rm s}\kappa_{\rm M}}^{(ij)}(\bm{\ell}),\\
P^{(ij)}_{mg}(\bm{\ell}) &=& 4(\alpha^{(i)}-1) \int_0^{\chi_H}d\chi\;\frac{q_{\rm M}^{(i)}(\chi)p_{\rm M}^{(j)}(\chi)}{f^2_K(\chi)}\nonumber\\
&& \times\; b\left(\frac{\bm{\ell}}{f_K(\chi)},\chi\right)P^{(ij)}_{\delta\delta}\left(\frac{\bm{\ell}}{f_K(\chi)},\chi\right),\\
P^{(ij)}_{gm}(\bm{\ell}) &=& P^{(ji)}_{mg}(\bm{\ell}), \\
P^{(ij)}_{gg}(\bm{\ell}) &=& \int_0^{\chi_H}d\chi\;\frac{p_{\rm M}^{(i)}(\chi)p_{\rm M}^{(j)}(\chi)}{f^2_K(\chi)}\nonumber\\
&&\times\;b^2\left(\frac{\bm{\ell}}{f_K(\chi)},\chi\right)P^{(ij)}_{\delta\delta}\left(\frac{\bm{\ell}}{f_K(\chi)},\chi\right),\\
P^{(ij)}_{\kappa_s g}(\bm{\ell}) &=& \int_0^{\chi_H}d\chi\;\frac{q_{\rm s}^{(i)}(\chi)p_{\rm M}^{(j)}(\chi)}{f^2_K(\chi)}\nonumber\\
&&\times\;b\left(\frac{\bm{\ell}}{f_K(\chi)},\chi\right)P^{(ij)}_{\delta\delta}\left(\frac{\bm{\ell}}{f_K(\chi)},\chi\right),\\
P^{(ij)}_{m \kappa_{\rm s}}(\bm{\ell}) &=& 2(\alpha^{(i)}-1)P^{(ij)}_{\kappa_{\rm M} \kappa_{\rm s}}(\bm{\ell}),
\eea
where
\be
P^{(ij)}_{\kappa_x\kappa_y}(\bm{\ell}) = \int_0^{\chi_H}d\chi\;\frac{q_x^{(i)}(\chi)q_y^{(j)}(\chi)}{f^2_K(\chi)}P^{(ij)}_{\delta\delta}\left(\frac{\bm{\ell}}{f_K(\chi)},\chi\right),
\ee
with $x,y = \{ {\rm s,M} \}$.

The final contribution to the observed number density contrast power spectrum comes from the shot noise term in Equation (\ref{eqn:NumberDensityContrastPS}), which takes the form
\be
P^{SN}_{\delta n} = \frac{1}{\langle n_{\delta n}\rangle^{(i)}},
\ee
where $\langle n_{\delta n} \rangle$ is the mean number density of sources used for the clustering analysis.

For ellipticity measurements, the shot noise is given by 
\be
P^{SN}_{\epsilon} = \frac{\sigma_\epsilon^2}{2\langle n_{\epsilon}\rangle^{(i)}},
\ee
where $\langle n_{\epsilon} \rangle $ is the mean number density of sources used in the shape analysis, and $\sigma_\epsilon$ is the total intrinsic ellipticity dispersion, taken here to be $\sigma_\epsilon = 0.4$. 

Subscripts on the global mean number density $\langle n\rangle$ account for the fact that the global number density of sources used for a shear analysis may differ to that using galaxy clustering information, due to different source redshift distributions and differing source samples, as a result of different selection techniques for the shape and photometry samples (for example, size cuts applied to the shape sample of galaxies).

\subsection{Parameter forecasts}\label{Sec:ParameterForecasts}

To estimate parameter constraints we use a Fisher Matrix analysis. We consider constraints for the set of cosmological parameters $Q =\{\Omega_M,\Omega_{{\rm Baryon}}, \Omega_{\Lambda}, w_0,  h, n_{{\rm s}}, \sigma_8\}$, which are the matter density, baryon density, dark energy density, dark energy equation of state, dimensionless Hubble parameter, spectral index giving the slope of the primordial power spectrum and the root--mean--square linear matter density fluctuations within a sphere of radius $8h^{-1}\rm{Mpc}^3$ respectively. These parameters are taken around fiducial values $\{0.3,0.0456,0.7,-1.0,0.7,1.0,0.8\}$. We do not restrict ourselves to flat models, and the curvature is set by $\Omega_k = 1-(\Omega_M + \Omega_{\Lambda})$. We choose $\ell_{\rm min} = 50$ to avoid inaccuracies in the Limber approximation, and let $\ell_{\rm max} = 5000$. However $\ell$-mode cuts are implemented on the clustering data to remove scales where the bias is expected to be non--linear (see Section \ref{Sec:GalaxyBiasModelling}).

The covariance between two power spectra is \citep{Joachimi:2010p855} 
\bea\label{Eqn:Cov2pt}
&{\rm Cov}[ P^{(ij)}_{\alpha \beta}(\ell),& P^{(rs)}_{\gamma \delta }(\ell')] = \delta_K^{\ell\ell'}\frac{2\pi}{\Delta\Omega\ell\Delta\ell}\times\\
&&\{P^{(ir)}_{\alpha\gamma}(\ell)P^{(js)}_{\beta\delta}(\ell)+P^{(is)}_{\alpha\delta}(\ell)P^{(jr)}_{\beta\gamma}(\ell)\}\nonumber\\
&& = {\rm Cov}_{\alpha\beta\gamma\delta}^{(ijrs)}(\bm{\ell}),
\eea
where subscripts $\alpha,\beta,\gamma,\delta = \{\delta n, \epsilon\}$, and we have assumed that the fields are Gaussian, an assumption that would need to be modified in a more sophisticated analysis. The Kronecker delta symbol marks the non--mixing of angular wavenumber modes, and $\Delta\Omega$ denotes the sky coverage area of the survey.

As the mean of the power spectrum is non--zero, the Fisher matrix is dominated by variations in the mean and given by \citep{Tegmark:1997p9}
\be\label{Eqn:FM2pt}
\mathbfss{F}_{\eta\tau} = \sum_\ell \bm{D}_{,\eta}(\bm{\ell}) \mathbfss{C}_{\rm A}^{-1}(\bm{\ell}) \bm{D}_{,\tau}(\bm{\ell})
\ee
where $\mathbfss{C}_{\rm A}(\bm{\ell})$ is the covariance matrix for data vector $\bm{D}(\bm{\ell})$ containing power spectra at angular wavenumber $\bm{\ell}_r$,  with mean of zero, notation `$,\eta$' means the partial derivative with respect to parameter $Q_{\eta}$. Subscripts $\eta$ and $\tau$ run over the set of cosmological parameters $Q$.

We consider various combinations of shear, clustering and galaxy--galaxy lensing analyses, listed in Table \ref{Table:Analysis_Types}. For the remainder of this paper, we use `clustering--only' (AutoCl, AllCl) to refer to an analysis which takes the number density contrast power spectra (Equation \ref{eqn:NumberDensityContrastPS}) as data, `shear--only' (Sh) to refer to an analysis which uses only the cosmic shear power spectra (Equation \ref{eqn:Galaxy-Galaxy_PS}) as data, and `galaxy--galaxy--lensing--only' (GGL) that which uses only the galaxy--galaxy lensing power spectra (Equation \ref{eqn:NumberDensityContrastPS-GGLensing}). For each analysis type, we construct a data vector:
\begin{enumerate}
\item{For the shear--only case (Sh), the data vector takes the form $\bm{D}(\bm{\ell})  = \bm{D}_{\rm s}(\bm{\ell}) = \{P_{\epsilon\epsilon}^{(11)}(\bm{\ell}), P_{\epsilon\epsilon}^{(12)}(\bm{\ell}), \ldots, P_{\epsilon\epsilon}^{(N_z N_z)}(\bm{\ell})\}$ and contains $N_z(N_z+1)/2$ shear power spectra for each $\ell$--mode, where $N_z$ is the number of redshift bins used in a tomographic analysis. The covariance matrix is given as $\mathbfss{C}_{\rm A}^{(ij)(rs)}(\bm{\ell}) = {\rm Cov}_{\epsilon\epsilon\epsilon\epsilon}^{(ijrs)}(\bm{\ell})$ for $i,j,r,s  = [0,N_z]$.}  
\item{When we consider the clustering--only case (AutoCl, AllCl), the data vector takes the form  $\bm{D}(\bm{\ell})  = \bm{D}_{\rm M}(\bm{\ell}) = \{P_{\delta n \delta n}^{(11)}(\bm{\ell}), P_{\delta n \delta n}^{(12)}(\bm{\ell}), \ldots, P_{\delta n \delta n}^{(N_z N_z)}(\bm{\ell})\}$ when using all redshift bin combinations,  and $\bm{D}(\bm{\ell}) = \bm{D}_{\rm M}(\bm{\ell}) = \{P_{\delta n \delta n}^{(11)}(\bm{\ell}), P_{\delta n \delta n}^{(22)}(\bm{\ell}), \ldots, P_{\delta n \delta n}^{(N_z N_z)}(\bm{\ell})\}$ when considering only auto correlations. The data vector then contains $N_z$ clustering power spectra for each $\ell$--mode when only auto correlations are considered, and $N_z(N_z+1)/2$ clustering power spectra when all redshift bin correlations are included. The covariance matrix then takes the form $\mathbfss{C}_{\rm A}^{(ij)(rs)}(\bm{\ell}) = {\rm Cov}_{\delta n\delta n\delta n\delta n}^{(ijrs)}(\bm{\ell})$ for $i,j,r,s  = [0,N_z]$.}
\item{For the galaxy--galaxy--lensing--only case (GGL), the data vector takes the form $\bm{D}(\bm{\ell})  = \bm{D}_{\rm GGL}(\bm{\ell}) = \{P_{\epsilon\delta n}^{(11)}(\bm{\ell}), P_{\epsilon\delta n}^{(12)}(\bm{\ell}),  P_{\epsilon\delta n}^{(21)}(\bm{\ell})  \ldots, P_{\epsilon\delta n}^{(N_z N_z)}(\bm{\ell})\}$ and contains $N_z^2$ power spectra for each $\ell$--mode. The covariance matrix is given as $\mathbfss{C}_{\rm A}^{(ij)(rs)}(\bm{\ell}) = {\rm Cov}_{\epsilon\delta n\epsilon\delta n}^{(ijrs)}(\bm{\ell})$ for $i,j,r,s  = [0,N_z]$.}  
\item{For the combination of clustering with shear (Sh+AutoCl,Sh+AllCl), we define the data vector as $\bm{D}(\bm{\ell}) = \{\bm{D}_{\rm M}(\bm{\ell}), \bm{D}_{\rm s}(\bm{\ell})\}$. The covariance matrix then takes block form
\be
\mathbfss{C}_{\rm A}(\bm{\ell}) = \left(\begin{array}{c|c} {\rm Cov}_{\delta n\delta n\delta n\delta n}^{(ijrs)}(\bm{\ell})& {\rm Cov}_{\delta n\delta n\epsilon\epsilon}^{(ijrs)}(\bm{\ell})\\\hline  {\rm Cov}_{\epsilon\epsilon\delta n\delta n}^{(ijrs)}(\bm{\ell})& {\rm Cov}_{\epsilon\epsilon\epsilon\epsilon}^{(ijrs)}(\bm{\ell})\end{array}\right).
\ee
}
\item{Finally for the combination of clustering with shear and galaxy--galaxy lensing (Sh+AutoCl+GGL,Sh+AllCl+GGL), we define the data vector as $\bm{D}(\bm{\ell}) = \{\bm{D}_{\rm M}(\bm{\ell}), \bm{D}_{\rm GGL}(\bm{\ell}), \bm{D}_{\rm s}(\bm{\ell})\}$. The covariance matrix then takes block form
\be
\mathbfss{C}_{\rm A}(\bm{\ell}) = \left(\begin{array}{c|c|c} {\rm Cov}_{\delta n\delta n\delta n\delta n}^{(ijrs)}(\bm{\ell})& {\rm Cov}_{\delta n\delta n\delta n\epsilon}(\bm{\ell}) &{\rm Cov}_{\delta n\delta n\epsilon\epsilon}^{(ijrs)}(\bm{\ell})\\\hline
{\rm Cov}_{\delta n\epsilon\delta n\delta n}^{(ijrs)}(\bm{\ell})& {\rm Cov}_{\delta n\epsilon\delta n\epsilon}^{(ijrs)}(\bm{\ell}) & {\rm Cov}_{\delta n\epsilon\epsilon\epsilon}^{(ijrs)}(\bm{\ell})\\\hline
{\rm Cov}_{\epsilon\epsilon\delta n\delta n}^{(ijrs)}(\bm{\ell})& {\rm Cov}_{\epsilon\epsilon\delta n\epsilon}^{(ijrs)}(\bm{\ell}) & {\rm Cov}_{\epsilon\epsilon\epsilon\epsilon}^{(ijrs)}(\bm{\ell})\end{array}\right).
\ee
}
\end{enumerate}

\begin{table}
\begin{center}
\caption{List of all analyses considered here, with power spectra which enter the data vector, listed in Equations \ref{eqn:NumberDensityContrastPS} through \ref{eqn:NumberDensityContrastPS-GGLensing}. Throughout, `Sh' labels cosmic shear, `GGL' labels galaxy--galaxy lensing, `AllCl' labels a clustering analysis using all redshift bin combinations, whilst `AutoCl' labels a clustering analysis using only auto--correlations in redshift.}\label{Table:Analysis_Types}
{
\renewcommand{\arraystretch}{1.8}
\begin{tabular}[]{| l | l |}
\hline
Analysis Type & Data \\
\hline
AutoCl  & 				$\delta_K^{ij}P_{\delta n \delta n}^{(ij)}$				 \\
AllCl  & 				$P_{\delta n \delta n}^{(ij)}$				 \\
GGL &				$P^{(ij)}_{\epsilon\delta n}$				 \\
AutoCl + GGL &			$\delta_K^{ij}P_{\delta n \delta n}^{(ij)}$, $P^{(ij)}_{\epsilon\delta n}$				 \\
AllCl + GGL &			$P_{\delta n \delta n}^{(ij)}$, $P^{(ij)}_{\epsilon\delta n}$				 \\
Sh  & 				$P^{(ij)}_{\epsilon\epsilon}$				\\
Sh + GGL  &			$P^{(ij)}_{\epsilon\epsilon}$, $P^{(ij)}_{\epsilon\delta n}$				 \\
Sh+ AutoCl & 			$P^{(ij)}_{\epsilon\epsilon}$, $\delta_K^{ij}P_{\delta n \delta n}^{(ij)}$				\\
Sh + AllCl  & 			$P^{(ij)}_{\epsilon\epsilon}$, $P_{\delta n \delta n}^{(ij)}$				\\
Sh + AutoCl + GGL &		$P^{(ij)}_{\epsilon\epsilon}$, $\delta_K^{ij}P_{\delta n \delta n}^{(ij)}$, $P^{(ij)}_{\epsilon\delta n}$				\\
Sh + AllCl + GGL & 		$P^{(ij)}_{\epsilon\epsilon}$, $P_{\delta n \delta n}^{(ij)}$,	 $P^{(ij)}_{\epsilon\delta n}$			\\
\hline
\end{tabular}
}
\end{center}
\end{table}

Throughout this paper we utilise a figure of merit (FoM) as a measure of the constraining power of either analysis considered above, defined as
\be
{\rm FoM} = {{{\rm det}([\mathbfss{F}^{-1}]_q)}}^{-\frac{1}{n_q}},
\ee
where $q$ denotes the subset of parameter space we are interested in, and $n_q$ the number of parameters in that subset, so that the figure of merit has been rescaled to one dimension. In this paper we consider two types of FoM:

\begin{enumerate}
\item{(FoM$_{\rm DE}$) A Dark Energy Task Force--like figure of merit \citep{Albrecht:2006}, taking $q=\{\Omega_{\Lambda},w\}$.}
\item{(FoM$_{\rm Cos}$) taking $q=Q$ to be the full cosmology parameter set, and thus containing information on the constraints on all parameters.}
\end{enumerate}

\section{Modelling}\label{Sec:Modelling}

\subsection{Galaxy bias}\label{Sec:GalaxyBiasModelling}

In Section \ref{Sec:2ptCorrelations} we discussed briefly our parameterisation of the intrinsic clustering correlations using a galaxy bias parameter which can be both scale and redshift dependent, $b(k,z)$. We discard any information from the regime where the galaxy bias is expected to be non--linear, and least well--known. To do this we utilise a similar technique to that in \cite{Joachimi:2010p855} and \cite{Rassat:2008p2019}, and remove information at the Fisher matrix level by discarding any bias--dependent information above a certain $\ell$--cut ($\ell_{\rm max}$), where
\be
\ell_{\rm max}(z^{(i)}) = f_K[\chi(z^{(i)})]k_{\rm max}(z^{(i)}_{\rm max}).
\ee

We choose the median redshift as the characteristic redshift for bin $(i)$, $z^{(i)}$. The maximum wavenumber is fit as $k_{\rm max} = 1.4\pi/R_{\rm max}$, where $R_{\rm max}$ is the radius beyond which the r.m.s. variations in the matter over--density fall below a certain value:
\be
\sigma^2(R_{\rm max},z) = \int\;d\;\ln(k) \Delta^2(k,z)W^2(kR) = \sigma_R^2,
\ee
using the Fourier transform of a spherical top--hat window function $W(y) = [\sin(y) - y\cos(y)]/y^3$, and $kR = 1.4\pi$ is the value at which this window function first crosses zero, and beyond which there is negligible contribution to the variance from the dark matter power. Hereafter we choose to take $\ell$--cuts by fitting $k_{\rm max}$ such that the matter density variance with a sphere of radius $R$ is $\sigma(R,z) \le 0.5$, which gives $k_{\rm max} \approx 0.3 \;h{\rm Mpc}^{-1}$ at $z = 0$, corresponding to $R \approx 14\; h^{-1}{\rm Mpc}$, within the quasi--linear regime. We investigate the implications of this choice of $\sigma_R$ in Section \ref{Sec:ell-cut-dependance}.  

For the number density contrast power spectra $P^{(ij)}_{\delta n\delta n}$, we take the $\ell$--cut corresponding to the lowest redshift bin, and for the shear--number density contrast power spectra $P^{(ij)}_{\epsilon \delta n}$ we impose $\ell$--cuts for the redshift bin from which number density information is obtained (in this example $\ell^{(j)}_{\rm max}$).

We assume the galaxy bias is scale--independent, and model the redshift dependence using one nuisance parameter per redshift bin, each varying independently and without bound. We consider three scenarios: Firstly, all galaxy bias nuisance parameters are held fixed at their fiducial value, taken to be $b_{\rm fid}^{(i)} = 1$ for all redshift bins. This corresponds to the case where the linear galaxy bias is perfectly known. Secondly, all galaxy bias nuisance parameters are constrained by the data, with no prior.

In a third scenario we add a prior on the galaxy bias, where the covariance matrix of the galaxy bias parameters is modelled to take the form:
\be\label{Eqn:BiasPriorCov}
\mathbfss{C}_{{\rm Bias}} = \sigma_\nu^2 \begin{pmatrix} 1 & \nu & \nu^2  & \cdots & \nu^{N_z-1}\\ \nu & 1 & \nu & \cdots & \nu^{N_z-2} \\  \vdots &  & \ddots & & \vdots \\  \nu^{N_z-1} & \nu^{N_z-2} & \nu^{N_z-3} & \cdots & 1\end{pmatrix}.
\ee
where $\nu$ gives the strength of the correlation between adjacent bins, and $\sigma_\nu$ is 
\be\label{eqn:BP_sigr_sig0}
\sigma_\nu = \sigma_0\left[\frac{N_z-(N_z-2)\nu}{N_z(1+\nu)}\right]^{\frac{1}{2}}.
\ee
to ensure 1--$\sigma$ errors on each bias parameter along $b_1 = b_2=  \ldots = b_{N_z} \equiv b$ are independent of $\nu$ (see Appendix \ref{App:NormalisingGBP}), and where $\sigma_0$ is the uncertainty of each bias parameter. This prior is then added to the Fisher matrix as 
\be
\mathbfss{F}_{\eta\tau} \to \mathbfss{F}_{\eta\tau} + (\mathbfss{C}^{-1}_{\rm Bias})_{{\eta\tau}}
\ee
where in this case $\eta$ and $\tau$ run only over the bias parameters, $\eta,\tau = \{b^{(1)},\cdots,b^{(N_z)}\}$, and all elements corresponding to all other parameters are left unchanged.

 We therefore define a correlation length, equivalent to the redshift over which the adjacent galaxy bias parameters are significantly correlated. By increasing the strength of the correlation, we reduce the freedom each bias parameter has with respect to its neighbour, thus reducing the variance of the galaxy bias nuisance parameters across redshift bins and making the function of bias versus redshift smoother. We therefore expect that as we increase the correlation strength ($\nu\to1$), we should improve cosmological constraints from clustering measurements as we retain more information from the clustering of sources, and increase the value of a FoM using cosmological parameters. As we expect galaxy bias to be a smooth function in redshift, we do not expect galaxy bias parameters between adjacent redshift bins to be negatively correlated, and we consider only $0\le \nu \le 1$. It is worth noting that with the galaxy bias prior defined in Equation (\ref{eqn:BP_sigr_sig0}), the prior becomes singular as $\nu\to 1$.

 The parameter $\sigma_0$ gives the marginal error on each bias nuisance parameter when fully uncorrelated ($\nu=0$). As such, $\sigma_0$ gives the scatter on the value of the bias in each redshift bin, and is assumed to be the same across all redshift bins considered. As $\sigma_0$ sets the level of uncertainty on each galaxy bias parameter, we expect that as we increase the uncertainty in galaxy bias, the constraining power from galaxy clustering will be reduced as less information from the clustering signal is accessible (conversely as $\sigma_0\to 0$, we recover fully known galaxy bias).

\subsection{The slope of the galaxy luminosity function}\label{Sec:CFHT}

In choosing the values for the slope of the number counts ($\alpha$) which sets the strength of the magnification effect, we first investigate what typical $\alpha$ values we would expect for an optical galaxy sample using public catalogues from the Canada--France--Hawaii Lensing Survey (CFHTLenS).

This survey covers 154 deg$^2$, ($\sim125$ deg$^2$ after masking) of the sky in the five $u^*g'r'i'z'$ filters. CFHTLenS combines weak lensing data processed with THELI \citep{Erben:2012p1556}, shear measurement using lensfit \citep{Miller:2013p1648} and photometric redshift measurement using PSF--matched photometry \citep{Hildebrandt:2012p2016}, with full systematic error analysis of shear measurements and photometry in \cite{Heymans:2012p1653}, and further error analysis of photometric redshifts in \cite{Benjamin:2013p2017}. 

\begin{figure}
\centering
\includegraphics[width = 0.5\textwidth]{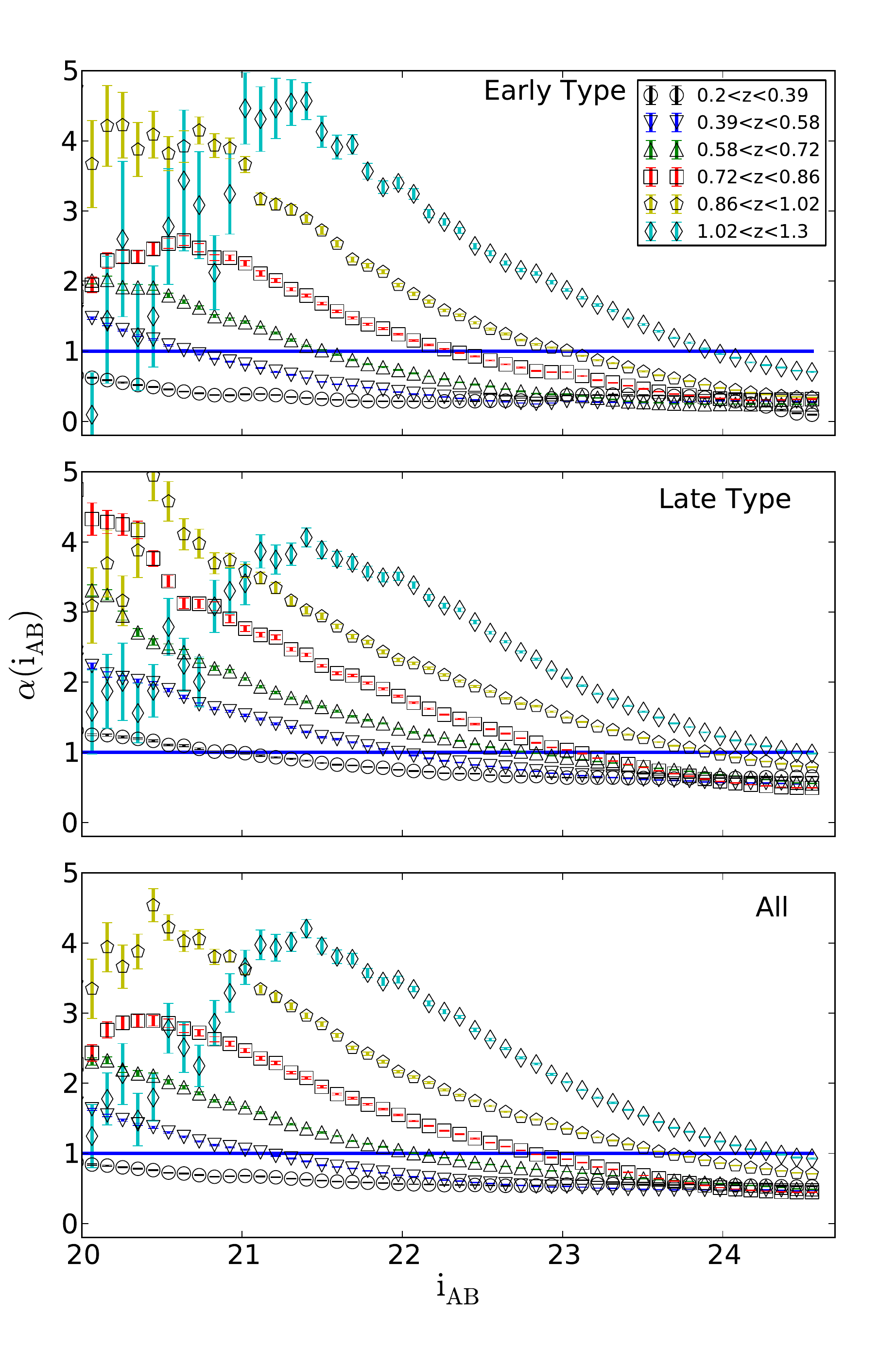}
\caption{The slope of the cumulative galaxy number counts, $\alpha$,  as a function of magnitude for a sample of redshift bins, chosen to be the set of tomographic bins used in \citet{Heymans:2013p2015}, for a population consisting of early--type galaxies (upper), late--type galaxies (middle) and the full combined sample (lower). } \label{fig:Alpha_Mag_Redshift}
\end{figure}

We construct the cumulative number densities of galaxies as a function of limiting magnitude, where sources have been separated into redshift bins, and take the slope of the cumulative number counts as set out in Equation (\ref{Eqn:Alpha_Magnitude}) to calculate $\alpha$.  Fig. \ref{fig:Alpha_Mag_Redshift} shows $\alpha$ as a function of magnitude limit for six tomographic redshift bins as used in \cite{Heymans:2013p2015} for three sets of galaxy samples split by population type according to their spectral energy distribution (SED) template value ($T$): late--type galaxies characterised by $T\ge2$, early--types taken to have $T\le 2$, and a full sample (labelled ``All''). Errors are taken from a bootstrap analysis, and agree with the estimated errors by taking the variance of $\alpha$ across all fields. 

Whilst the value of $\alpha$ varies across the range of limiting magnitudes, all redshift bins converge to a similar value near the limiting magnitude of CFHTLenS at $i_{\rm AB} = 24.7$. The maximum value of $\alpha$ may occur at lower magnitudes. However by cutting at lower magnitudes we reduce the number of galaxies in the sample thus increasing the shot noise contribution. This suggests that choosing a magnitude cut to maximise $\alpha$ (and consequently the strength of the contribution from cosmic magnification to the clustering signal) may not maximise the signal--to--noise ratio for a flux magnification analysis. 

\begin{figure}
\centering
\includegraphics[width = 0.5\textwidth]{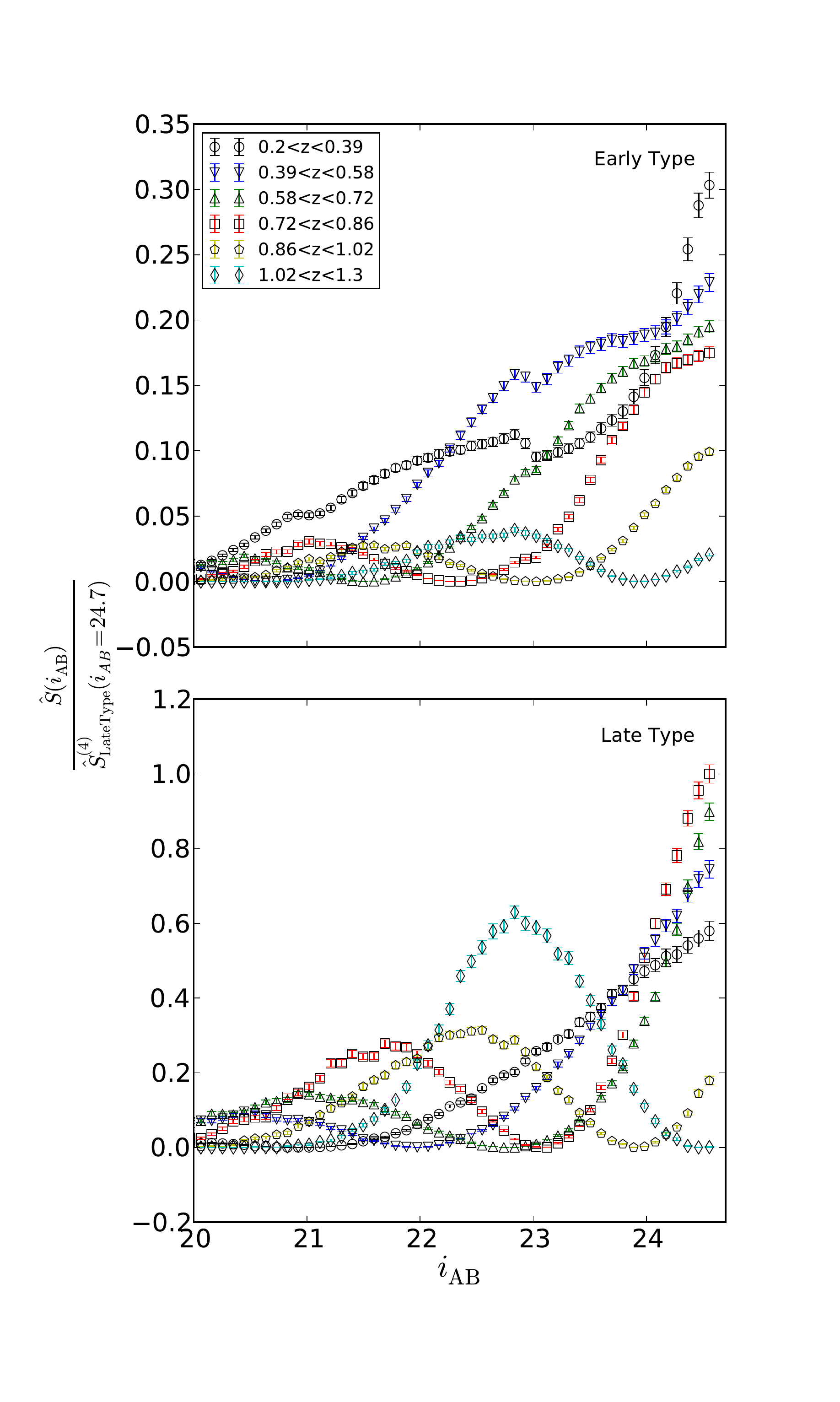}
\caption{The signal--to--noise ratio estimate $\hat{S}$ as a function of magnitude for the set of tomographic bins used in \citet{Heymans:2013p2015}, for a population consisting of early-- and late--type galaxies. To aid comparison, signal--to--noise ratio values are rescaled to the maximum signal--to--noise ratio value at the magnitude limit of the survey, which occurs in redshift bin $4$, $0.72<z<0.86$ in the late--type subsample. The signal--to--noise ratio for the late--type subsample is typically larger than the early--type subsample, which can be attributed to the smaller early--type population in the full catalogues. Note the different range of ordinates.}\label{Fig:SignalToNoise}
\end{figure}

We therefore define a signal--to--noise ratio estimate for background redshift bin $(j)$ lensed by foreground redshift bin $(1)$ as
\be\label{eqn:SignalToNoiseEst}
\hat{S}^{(j)}(i_{\rm AB}) = \langle n^{(1)} \rangle \langle n^{(j)}(i_{\rm AB})\rangle[\alpha^{(j)}(i_{\rm AB})-1]^2,
\ee
where the magnitude limit for the foreground bin is chosen to be as deep as possible to maximise $\langle n^{(1)}\rangle$, and is proportional to the square of the true signal--to--noise ratio. To construct this signal--to--noise ratio estimator $\hat{S}$ we have considered the case where the  magnification--intrinsic clustering power spectrum $P_{gm}$ is the only contribution to the clustering signal, and cosmic variance has been assumed to be negligible. Fig. \ref{Fig:SignalToNoise} shows how $\hat{S}$, rescaled by the maximum signal--to--noise ratio at the limiting magnitude of the survey, behaves as a function of limiting $i$ magnitude, for early and late--type galaxy subsamples. 

From Fig. \ref{fig:Alpha_Mag_Redshift} it is noticeable that the qualitative variation of $\alpha$ as a function of magnitude and redshift is similar for both samples of late-- and early--type galaxies, however the value of $\alpha$ at a given magnitude and redshift is typically larger for the late--type sample.
As the catalogues used contain predominantly late--type galaxies, the behaviour of $\alpha$ as a function of magnitude in the full sample is set mainly by the late--type galaxies, and similarly for the signal--to--noise ratio estimator. 
Whilst both samples show similar qualitative variation in $\alpha$ as a function of magnitude and redshift, the difference between the two samples is more marked in their respective signal--to--noise plots.
Typically, the signal--to--noise ratio for the late--type galaxies is larger than those for the early--type subsample, and this can be attributed to the fact that the late--type subsample contains more galaxies thus reducing the shot noise contribution. 

For the late--type subsample, all redshift bins show an oscillation in the signal--to--noise ratio as the magnitude is varied. For all redshift bins, we see an increase in the signal--to--noise ratio as the noise is reduced by increasing the size of the sample, followed by a reduction in the signal--to--noise ratio as $\alpha\to 1$, the limit where there is no change in the number density of sources due to flux magnification. Finally, past this limit and for $\alpha < 1$ the reduction in noise from increased sample size causes an increase in the signal--to--noise ratio. 

For the redshift bins encompassing $0.2\le z<0.86$ the peak of the signal--to--noise ratio estimator at the faint limit of the survey ($i_{\rm AB} = 24.7$) is larger than the peak of the oscillation for each respective redshift bin. However in the two largest redshifts bins, the peaks of these oscillations give a larger signal--to--noise ratio at a brighter magnitude than that obtained by naively cutting at the faint limit of the survey. This suggests an analysis chosen to optimise the strength of the magnification contribution to the measured clustering power spectrum should cut at lower magnitude limits for these redshift bins. However, peaks in the signal--to--noise estimator at magnitudes lower than the limit of the survey are limited to the highest redshift bins, where photometry is least accurate, and are nearly entirely absent in the early--type subsample.

In this analysis, we have chosen to consider the optimisation of the clustering signal as set out in Equation \ref{eqn:NumberDensityContrastPS}, however the signal--to--noise of the flux magnification contribution may also be optimised by weighting the sample according to $(\alpha - 1)$ \citep{MenardBartelmann2002}. We leave an analysis using this optimisation for future work.

As well as adding galaxies at the faint end of the luminosity function, magnification should also remove galaxies at the bright end of the luminosity function. Assuming that all galaxies in a redshift bin experience a change in magnitude $\Delta m$ as a result of lensing by the foreground, then the change in the number of galaxies at the faint end is roughly $N(m_{\rm faint})\Delta m$, whilst the change at the bright end is $\sim N(m_{\rm bright})\Delta m$ \footnote{This ignores the second--order effect of the change in the cumulative number of sources brighter than the faint end by the removal of sources at the bright end}. Since $N(m_{\rm bright})/N(m_{\rm faint})\ll1$, we expect the effect of the removal of sources at the bright end to be subdominant, and ignore it. By choosing a magnitude cut close to the bright limit, this effect becomes more important. 

For these reasons, we limit the choice of $\alpha$ values only to values close to the faint limit of the survey, and consider $0<\alpha<4$ as reasonable. Unless otherwise stated, results are shown assuming $\alpha = \ATrue$ across all redshift bins, which is the value calculated for the full galaxy sample for $i_{\rm AB}\le 24.7$ and $z\le 0.86$.

\subsection{Survey modelling}\label{Sec:SurveyModelling}

In this analysis, we consider two types of survey, following a Dark Energy Task Force--like classification of surveys \citep{Albrecht:2006} considering
\begin{enumerate}
\item{A Stage III ground--based survey (hereafter S3), covering $1,500\:{\rm deg}^2$ to a depth of $i_{\rm AB} = 24$. Sources are measured between photometric redshift limits  $z_{\rm Phot} = (0,2)$ with photometric redshift errors $\sigma_{\rm Phot} = 0.05(1+z_{\rm Phot})$}
\item{A Stage IV space--based survey (hereafter S4), which covers $15,000 \:{\rm deg}^2$ of the sky, to a depth of 24.7 in the $i$--band, between photometric redshift limits $z_{\rm Phot} = (0,2)$, and with photometric redshift errors $\sigma_{\rm Phot} = 0.05(1+z_{\rm Phot})$. Unless otherwise stated, results hereafter are shown for an S4 survey.}
\end{enumerate}

\begin{figure}
\includegraphics[width = 0.5\textwidth, trim  =5mm 10mm 0mm 0mm]{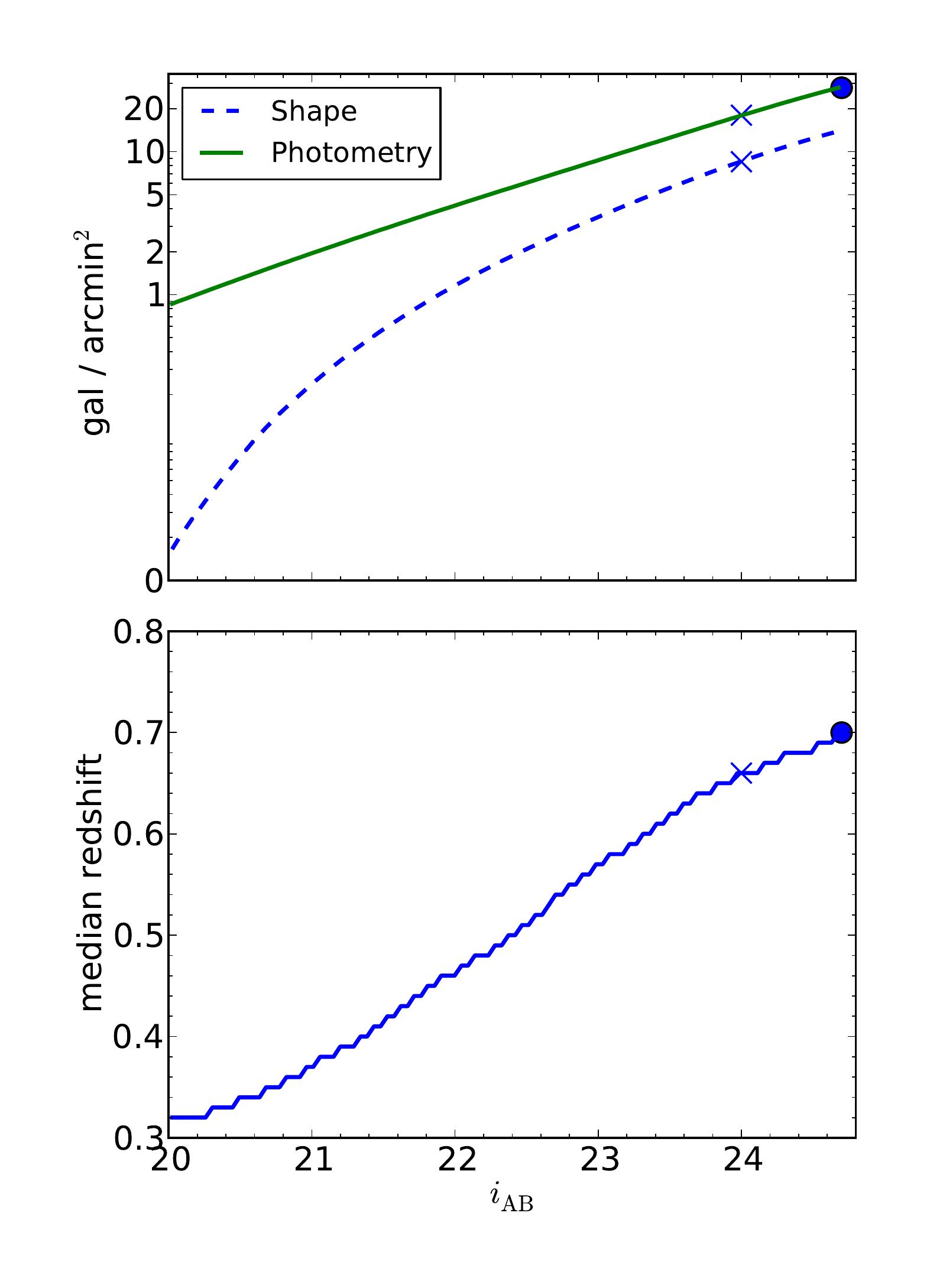}
\caption{The top panel shows the effective number density of galaxies (in galaxies/arcminute$^2$) as a function of the faint limiting magnitude $i_{\rm AB} $ for sources with valid shape measurement (dashed), and for all galaxies with photometry (solid). The bottom panel shows the median redshift of sources as a function of the limiting magnitude. All data is taken from CFHTLenS catalogues. Crosses mark values taken for an S3 survey, whilst dots mark values taken for an S4 survey.}\label{Fig:CFHT_Neff_Zmed}
\end{figure}

Fig. \ref{Fig:CFHT_Neff_Zmed} shows the median survey redshift, and effective number density of galaxies as a function of faint limiting $i_{\rm AB}$ magnitude taken from CFHTLenS catalogues for galaxies with valid shape measurement \citep{HeymansCFHTLenS} as well as for all galaxies with photometry. As the galaxies broad size distributions are weakly redshift dependent, cuts on galaxy size such as those used in a shape analysis do not significantly affect the measured median redshift, but will cause a noticeable decrease in effective number density of galaxies used. From Fig. \ref{Fig:CFHT_Neff_Zmed} we choose for our S3 survey a median redshift $z_{\rm med} = 0.66$, and galaxy number density of $\langle n\rangle = 8.5 \;{\rm galaxies/arcminute}^2$ and $18\;{\rm gal/arcminute}^2$ for shear and clustering analyses respectively.  Similarly, we consider for an S4 survey $z_{\rm med} = 0.7$, and assume shapes can be measured for every detected source. Therefore, we use the photometry line in deducing the number density of galaxies for S4, and take $n_{\rm gal} = 28\;{\rm galaxies/arcminute}^2$ for both shear and clustering analyses.  

\subsection{Galaxy redshift distributions}\label{Sec:Galaxy_Redshift_Distributions}

We model the distribution of galaxies with true redshift $z_{\rm t}$ as
\be
p^{(i)}(z_{\rm t}) = \int_{z^{(i)}_l}^{z^{(i)}_h} dz_{\rm ph}\; p(z_{\rm t}|z_{\rm ph})p(z_{\rm ph}),
\ee
where $p(z_{\rm t}|z_{\rm ph})$ is assumed to be a Gaussian with width $\sigma_z = 0.05(1+z_{\rm ph})$, and $z_l$ and $z_h$ denote the lower and higher bounds of the redshift bin in photometric redshift.  We model the galaxy photometric redshift distribution, $p(z_{\rm ph})$, as \citep{Smail:1994p986}
\be
p(z_{\rm ph}) \propto \left(\frac{z_{\rm ph}}{z_0}\right)^{2}e^{-\left(\frac{z_{\rm ph}}{z_0}\right)^{1.5}},
\ee
with characteristic redshift $z_0 = z_{\rm med}/1.412$. The galaxy redshift distribution is subdivided into redshift bins with equal numbers of galaxies. 

\begin{figure}
\includegraphics[width = 0.5\textwidth, trim  = 0mm 20mm 0mm 0mm]{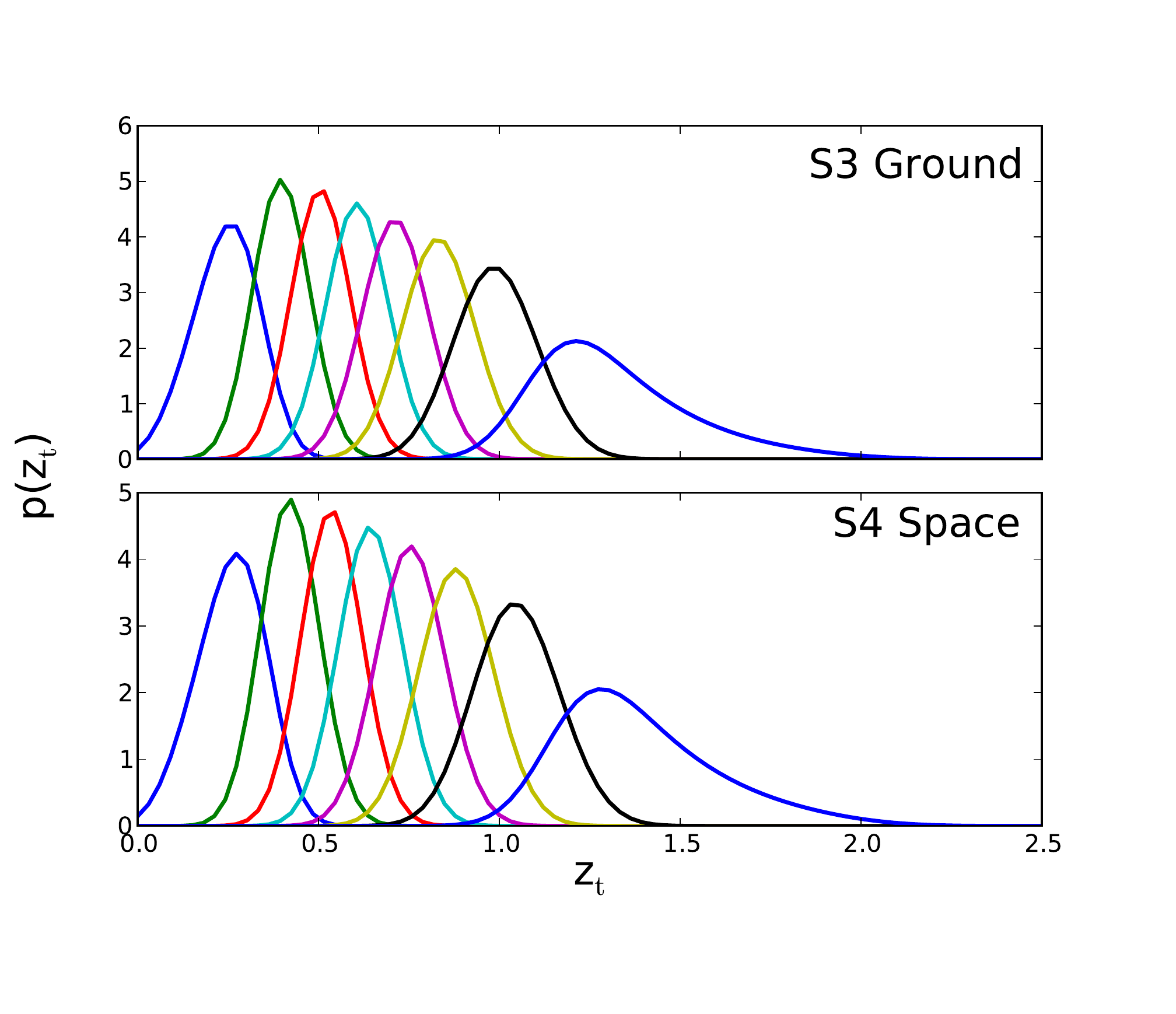}
\caption{Galaxy redshift probability distribution functions as defined in Section \ref{Sec:Galaxy_Redshift_Distributions}, for an S3 ground--based (top) and S4 space--based (bottom) survey (defined in Section \ref{Sec:SurveyModelling}).}\label{fig:SpecRedshiftDist}\label{Fig:Distributions}
\end{figure}

\begin{figure*}
\centering
\includegraphics[width = \textwidth]{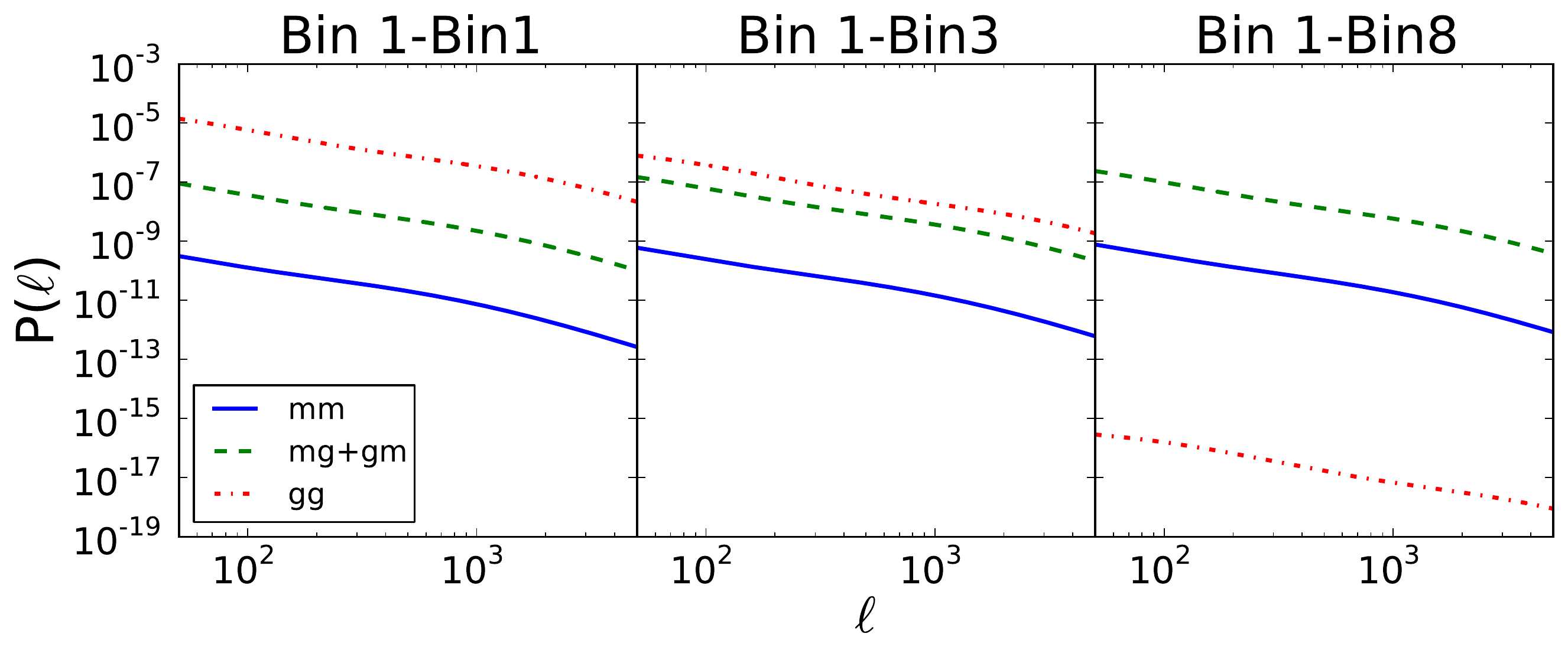}
\caption{Contributions to the number density contrast power spectrum for a combination of background redshift bins, for the S4 survey of Fig. \ref{Fig:Distributions}. The foreground bin is chosen to be redshift bin $1$. It can be seen that for foreground and background that are spatially close in redshift, the overlap in redshift distribution due to redshift errors can easily cause the magnification terms (mg or mm) to be swamped by the intrinsic clustering term (gg). As we increase the separation in redshift between foreground and background, the amplitude of the gg term decreases whilst the cross (mg + gm) and mm terms increase.} \label{Fig:NumberPowerSpectra}
\end{figure*}

Fig. \ref{Fig:Distributions} shows the resultant galaxy distribution for 8 redshift bins for survey types S3 and S4, with the number density contrast power spectra given for three redshift bin combinations in Fig. \ref{Fig:NumberPowerSpectra}, split by component. Noticeably, for closely--separated redshift bins, the intrinsic clustering term is non--vanishing due to the presence of photometric redshift errors which cause some galaxies to be incorrectly assigned to a given redshift bin, and causes overlap between the binned galaxy distributions. As we take redshift bins that are more widely separated, the power from intrinsic clustering decreases, so we see that for the most widely separated bins the total power is dominated by terms that include the magnification. It is for this reason that \cite{VanWaerbeke:2010p8} and \cite{Hildebrandt:2009p845} take spatially disjoint redshift bins to isolate the magnification contribution to the clustering power spectrum. As the cross power (mg+gm) is always dominant over the pure magnification (mm) power, frequently studies will ignore the pure magnification contribution to the overall clustering power, and instead just quote the cross contribution. In the situation where correlations are considered between distant foreground and backgrounds, the intrinsic clustering contribution is subdominant and may also be ignored. In this analysis, we consider all contributions to the power, as given in Equation \ref{eqn:NumberDensityContrastPS}.

\section{Results}\label{Sec:Results}

\subsection{The effect of galaxy bias}
\begin{figure*}
\centering
\includegraphics[width=\textwidth]{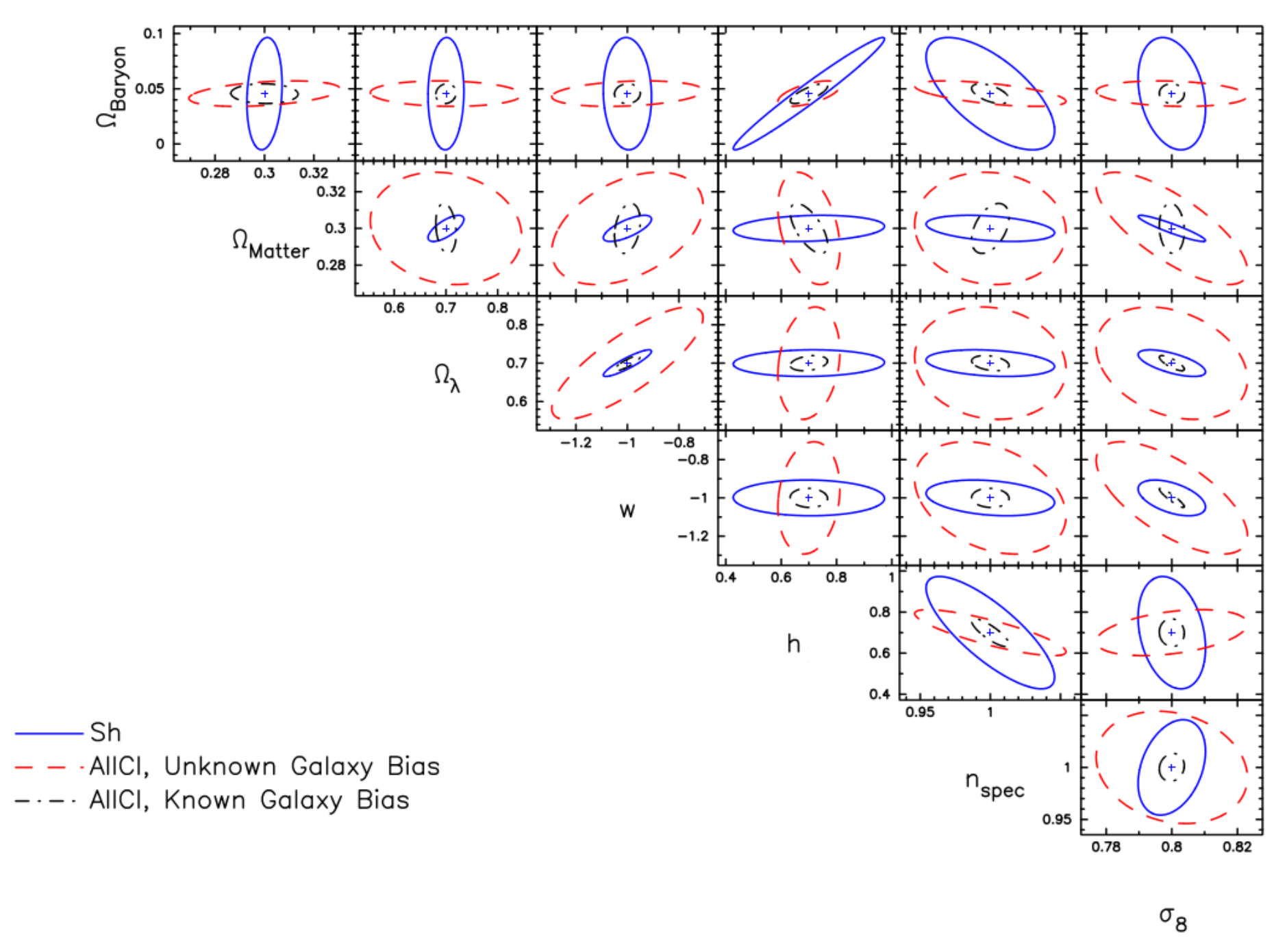}
\caption{Fisher matrix forecast showing marginal two--parameter, $1\sigma$ constraints for an S4 space--based survey, considering measurements of galaxy ellipticities only (``shear only'', solid blue line), and galaxy clustering including flux magnification. Fixed galaxy bias ($b=1$) is shown in black (dot--dashed), and unknown galaxy bias (simultaneously constrained with the data) is shown in red (dashed). Constraints from clustering with flux magnification assume $\alpha = \ATrue$, and only contain data from linear scales. Cuts on $\ell$--modes are applied as detailed in Section \ref{Sec:GalaxyBiasModelling}, with $\sigma_R < 0.5$. }\label{Fig:Contours_M_K}
\end{figure*}

\begin{figure*}
\centering
\includegraphics[width = \textwidth]{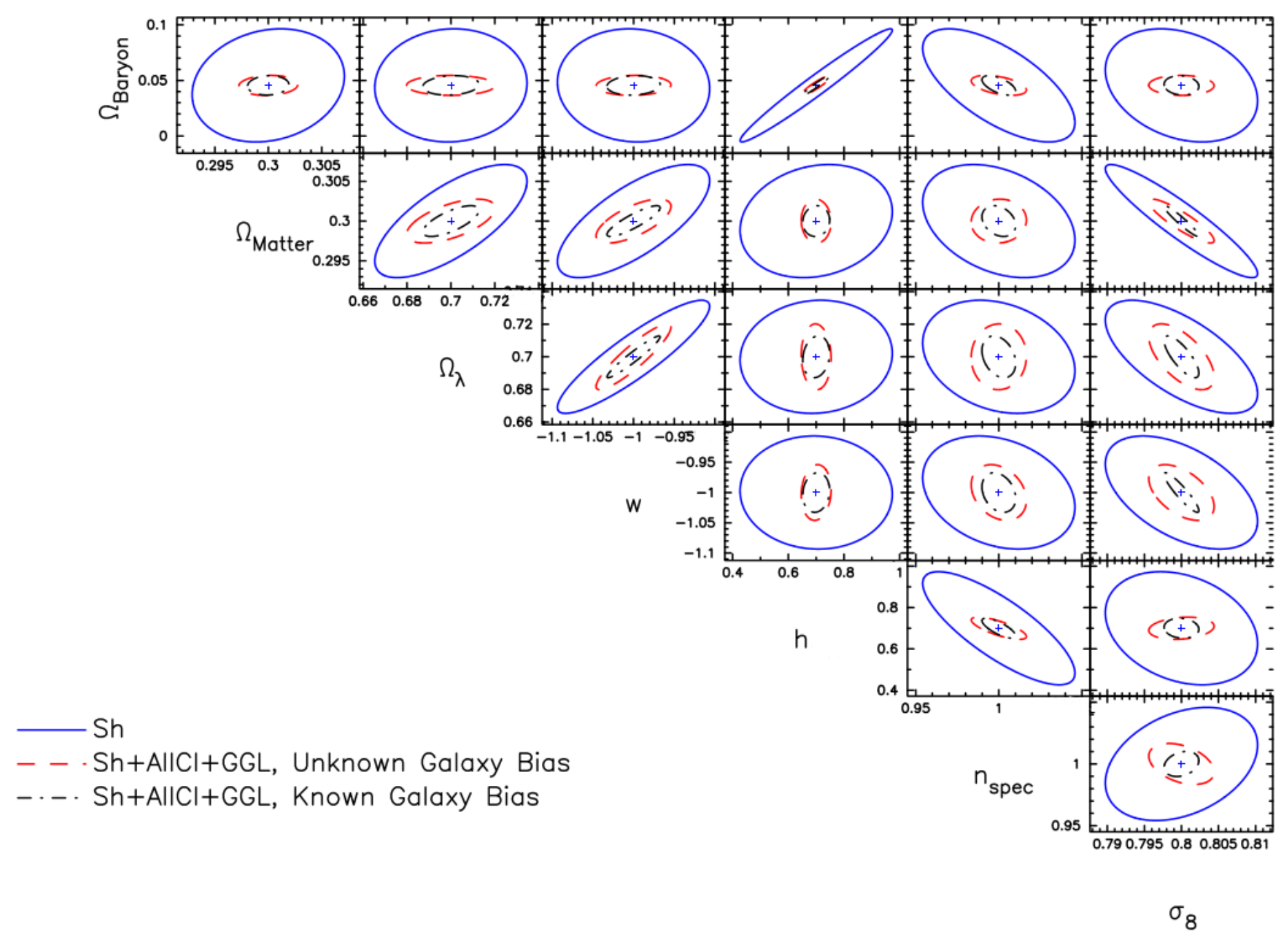}
\caption{As Fig. \ref{Fig:Contours_M_K}, but instead comparing measurements from galaxy ellipticities only (``shear only'', solid), with a combination of shear and galaxy clustering measurements including galaxy--galaxy lensing, for known galaxy bias, $b=1$, (dot--dashed), and unknown galaxy bias which is simultaneously constrained by the data (dashed).}\label{Fig:Contours_MK_K_MargFixed}
\end{figure*}

In this section, we present forecasts for our S3 and S4 survey models, using the Fisher matrix formalism set out in Section \ref{Sec:ParameterForecasts}. Throughout this section, we only consider a clustering analysis which includes all redshift bin correlations, and for which flux magnification is modelled (AllCl). Fig. \ref{Fig:Contours_M_K} shows contours for the set of cosmological parameters laid out in Section \ref{Sec:ParameterForecasts}, for the cases where we consider constraints coming from a shear--only (Sh), and clustering--only analysis using all redshift bin correlations (AllCl).  Fig. \ref{Fig:Contours_MK_K_MargFixed} shows $1\sigma$ parameter constraints for a shear--only analysis (Sh), and a combined shear and clustering analysis including galaxy--galaxy lensing (Sh+AllCl+GGL). Both figures show constraints only for a S4 survey, however results for both S3 and S4 surveys are summarised in Tables \ref{Table:FoM_DE_byAnalysis_S4} and \ref{Table:FoM_DE_byAnalysis_S3}. For analyses that contain number density contrast as a probe, $\ell$--cuts are applied and galaxy bias is either marginalised over (labelled as unknown galaxy bias), or fixed to $b = 1$ (labelled as known galaxy bias). 

It is evident that in the case where galaxy bias is known, constraints from a clustering analysis are competitive with cosmic shear. However when galaxy bias is unknown and must also be constrained from the data the constraints from clustering alone are much weaker. This is expected, as when the linear galaxy bias is known the intrinsic clustering contribution to the power spectrum directly probes the matter power spectrum. 

It is worth noting that constraints from clustering--only (AllCl) on $\Omega_{B}$ and $h$ are better than those from ellipticity measurements, as the clustering data can better pick out the turnover in the matter power spectrum , since the kernel for projected number density fluctuations (Equation \ref{eqn:Projected_NumberDensity}) is much narrower than that for cosmic shear (Equation \ref{eqn:ProjectedConvergence}). Whilst seemingly promising, distance probes combined with CMB measurements will also constrain these parameters very well.

For the  combined analysis (Sh+AllCl+GGL) there can be a significant improvement when adding clustering and galaxy--galaxy lensing to cosmic shear. However the improvement to constraints on cosmological parameters is dependent on whether galaxy bias is constrained using the data, or set to a fixed known value (Fig. \ref{Fig:Contours_MK_K_MargFixed}). If galaxy bias is known there is a marked improvement in constraints, especially in the $\Omega_M - \sigma_8$ plane. This is in agreement with the results of \cite{VanWaerbeke:2010p8}, in which galaxy bias was assumed linear and fixed to $b=1$. If galaxy bias is unknown some of the additional constraining power from clustering and galaxy--galaxy lensing is lost, with FoM$_{\rm DE}$ approximately $3.4$ times larger for the Sh+AllCl+GGL case if galaxy bias is fixed rather than free. We therefore conclude that the constraints presented in \cite{VanWaerbeke:2010p8} will be too optimistic. In fact, in the Sh+AllCl+GGL analysis, much of the information on bias comes from galaxy--galaxy lensing, which allows internal calibration of galaxy bias.  As a consequence, using both shear and number density to constrain both cosmology and galaxy bias simultaneously does not lead to much loss of information, provided the parameterisation of galaxy bias is realistic.

Even when the galaxy bias is unknown and simultaneously constrained with the clustering data, the improvement in parameter constraints from the addition of information from galaxy clustering and galaxy--galaxy lensing (Sh+AllCL+GGL) is significant, corresponding to an increase by a factor of $3.7$ in FoM$_{\rm DE}$ from its cosmic--shear--only value (Sh), for a S4 survey. We draw similar conclusions for the S3 model, with improvements in FoM value by a factor of $6.3$ for Sh+AllCl+GGL over the shear--only value. The larger improvement from a combined analysis over shear--only in this case is due to decrease in shot noise in the clustering correlations, as the photometric sample is larger than the shape sample for the S3 model.		

\begin{figure}
\centering
\includegraphics[width = 0.47\textwidth]{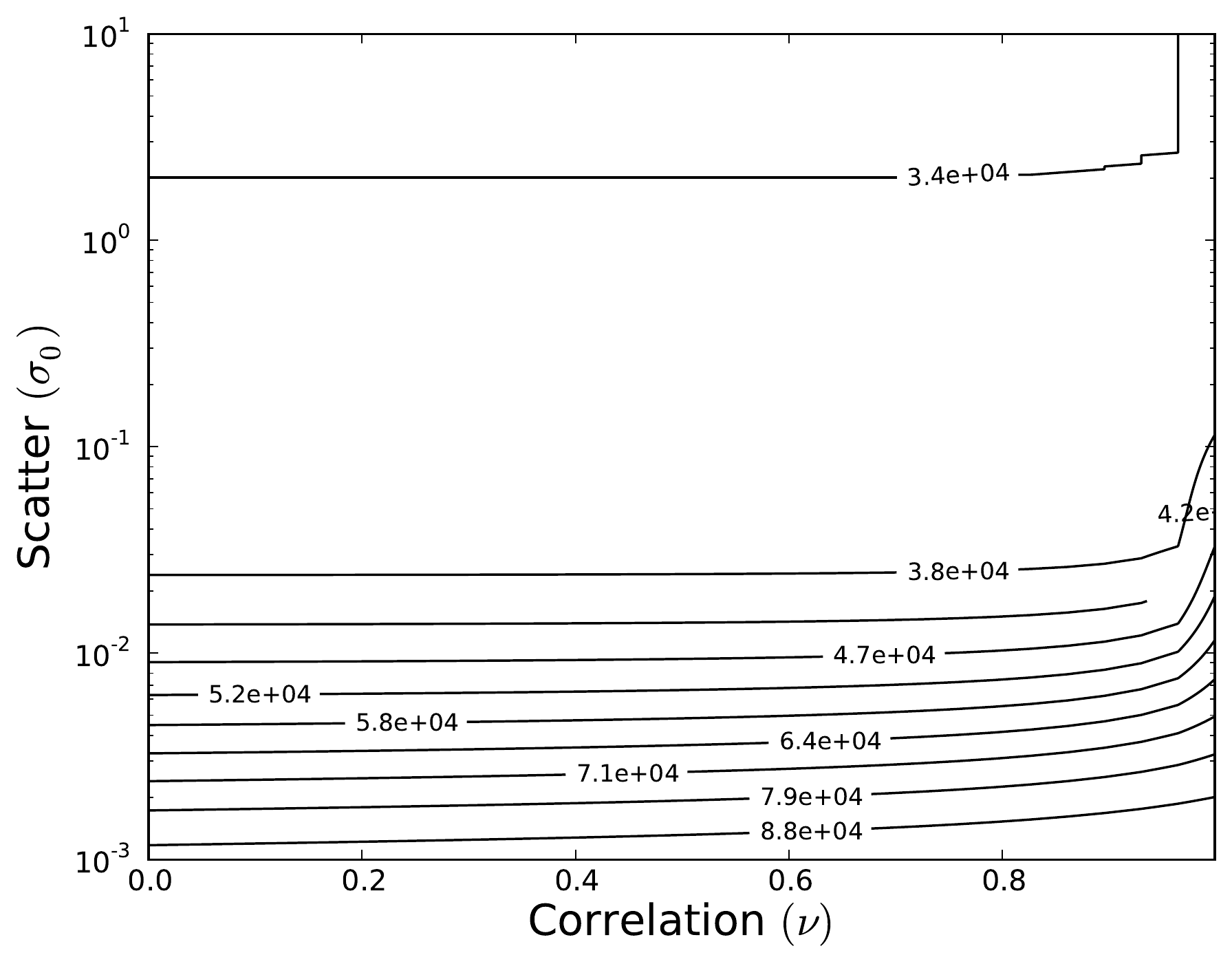}
\caption{FoM$_{\rm Cos}$ as a function of $\sigma_0$ and $\nu$ ( the uncertainty and correlation in the galaxy bias prior as detailed in Equation \ref{Eqn:BiasPriorCov}) for an S4 survey, using the Sh+AllCl+GGL analysis. }\label{Fig:BiasPrior_ContourPlot}
\end{figure}

If galaxy bias can be constrained externally, the picture will be intermediate between the scenarios presented so far. Therefore, we consider how much information can be regained by correlating galaxy bias parameters across redshift bins, equivalent to making the galaxy bias a smoother function in redshift, or limiting its uncertainty using an external probe, by the addition of a prior on galaxy bias of the form detailed in Section \ref{Sec:GalaxyBiasModelling}. Fig. \ref{Fig:BiasPrior_ContourPlot} shows $\rm{FoM_{Cos}}$ for a range of correlation strengths and uncertainties. It is evident that the uncertainty in galaxy bias affects the recovery of information. However there is only a weak degradation of figure of merit when the correlation strength is decreased for a joint (Sh+AllCl+GGL) analysis. 

\cite{Gaztanaga:2012p1194} concluded that magnification alone can produce better results than shear alone when the galaxy bias is known. However `magnification' there corresponds with `clustering' in this work, using all contributions to number density fluctuations. Accounting for the fact that we use different definitions for our figure of merits, our results are in broad agreement. There are differences between their analysis and that presented here: for example, they assume linear theory when modelling the matter power spectrum, and apply $\ell$--cuts to all probes, including their shear measurements, suggesting they under--estimate the constraining power of cosmic shear. Additionally, they modelled galaxy bias using four free parameters, whereas we assign a galaxy bias nuisance parameter to each redshift bin used. 
As a result, \cite{Gaztanaga:2012p1194} found that clustering (or magnification in their terminology) with unknown galaxy bias is much more competitive with cosmic shear than we find in this work. However the conclusion in both analyses is that when galaxy bias is known, galaxy clustering can be a competitive probe of cosmology to cosmic shear alone.  Similarly, both analyses show that the combination of galaxy clustering, galaxy--galaxy lensing and cosmic shear gives a significant improvement in statistical errors on cosmological parameters over cosmic shear alone.

\cite{EiflerKrause2013} consider a non--tomographic analysis which includes all cross--correlations between a intrinsic clustering, shear, galaxy--galaxy lensing, and magnification for a survey modelled on DES. The authors conclude that whilst the combination of all probes significantly improves the constraints over each individually, the inclusion of a magnification analysis does not substantially contribute to the information in joint shear, clustering and galaxy--galaxy lensing analysis, in agreement with the results we present here. However, the authors go beyond the assumption that the data are Gaussian distributed, and model a non--Gaussian contribution by summing tri--spectrum contributions using the halo model. They conclude that the change in forecasting cosmological parameter constraints for the full combined analysis due to incorrectly assuming Gaussianity is comparable to the change in constraints between taking known or free galaxy bias nuisance parameters. This change is largest for $\sigma_8$ and $n_s$, with the dark energy equation of state $w$ insensitive to non--Gaussianity. However there is no cut on highly non--linear scales in the clustering signal, and as we expect that the effect would be reduced if highly non--linear scales are removed from the analysis, we therefore expect our results to be robust even under the Gaussian assumption, particularly for dark energy parameters. 

\subsection{The contribution from flux magnification}\label{Sec:Constraints_AlphaDependance}

In this section, we investigate how much of the constraining power using galaxy clustering (AllCl) comes from the magnification terms ($P_{mm}, P_{mg}+P_{gm} $ in Equation \ref{eqn:NumberDensityContrastPS}), and how much comes from intrinsic clustering only ($P_{gg}$). If $\alpha = 1$, the terms which depend on the magnification are identically zero, so that number density fluctuations come from the intrinsic clustering only. In the limit of large $\alpha$, the clustering power spectrum is dominated by the magnification contribution ($P_{mm}$) for all redshift bin combinations. By altering $\alpha$, we can therefore alter the strength of the contribution from flux magnification, and therefore test the level of contribution to parameter constraints from the magnification effect.

\begin{figure}
\centering
\includegraphics[width = 0.5\textwidth, trim  = 0mm 10mm 3mm 5mm]{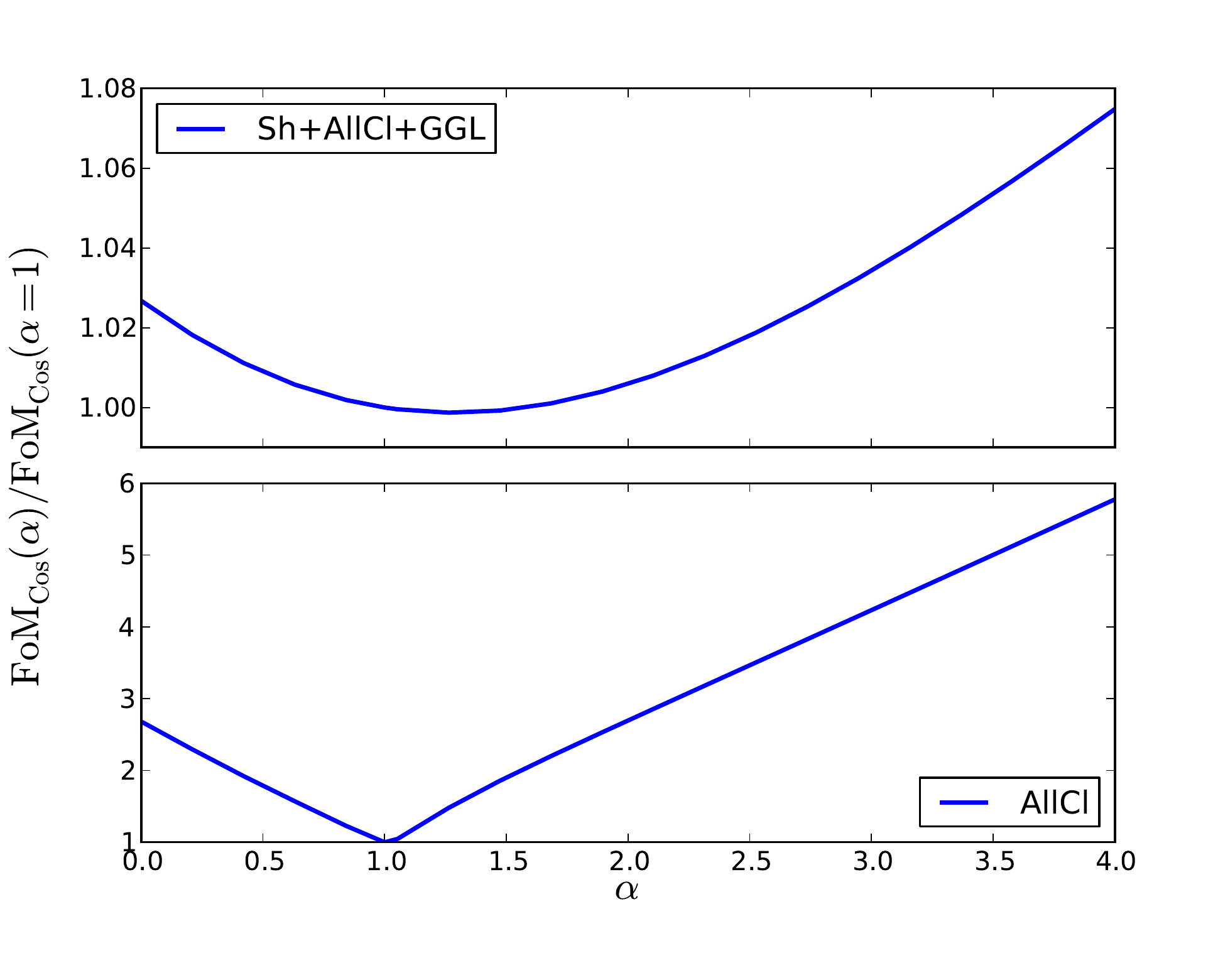}
\caption{Figure of Merit (FoM$_{\rm Cos}$) as a function of the slope of the cumulative number counts, $\alpha$, for a clustering--only analysis (bottom), and combined shear and clustering analysis (top), for an S4 survey. All values are rescaled to the value of the Figure of Merit when $\alpha = 1$ corresponding to the case when fluctuations to the number density contrast come from the intrinsic clustering of galaxies only. Note the different range of the ordinate axes.}\label{Fig:FoM_Alpha}
\end{figure}

Fig. \ref{Fig:FoM_Alpha} shows the figure of merit (FoM$_{\rm Cos}$) as a function of $\alpha$ from galaxy clustering--only (AllCl), and a combined clustering, cosmic shear and galaxy--galaxy lensing (Sh+AllCl+GGL) analysis. We note that the FoM increases with $\alpha>1$ for the AllCl case, as the contribution from flux magnification becomes larger, and the minimum in FoM$_{\rm Cos}$ occurs at $\alpha = 1$ where the contribution to the clustering power spectrum from the magnification terms is zero. We can therefore conclude that a non--zero measurement of $\alpha-1$ will improve the total constraints on cosmological parameters provided no galaxies are removed from the sample. However, as discussed in Section \ref{Sec:CFHT}, analyses that try to optimise for a clustering analysis by cutting at a magnitude limit which gives a large value of $\alpha$ may degrade the signal overall by removing sources from the sample and increasing noise.

Whilst similar behaviour is seen in the combined (Sh+AllCl+GGL) probe, we note that the relative improvement in the FoM as $|\alpha -1|>0$ is much smaller than the clustering--only (AllCl) case. As the constraints on cosmological parameters from clustering are much weaker that those from shear when galaxy bias is unknown, most of the improvement in FoM for the combined Sh+AllCl+GGL probe comes from degeneracy lifting between the clustering, shear and galaxy--galaxy lensing signals, and so, whilst a non--zero value of $\alpha-1$ may greatly improve the constraining power of a clustering--only (AllCl) analysis, it does not change the constraining power of a combined (Sh+AllCl+GGL) analysis significantly. We note also that the minimum of the FoM for the combined analysis does not lie at $\alpha  = 1$ as it does in the clustering--only case. For $\alpha>1$, the clustering signal becomes more degenerate with the shear, and so we expect a shift in the minimum for the combined analysis. Over this minimum, the increase in $\alpha$ causes the FoM for the combined analysis to increase.

In this initial study, we have chosen to ignore intrinsic alignments, for improved clarity of the magnification effects. The analysis here would therefore be most applicable for blue samples, for which the intrinsic alignment is small \citep{Heymans:2013p2015}. For a sample of red galaxies, a more comprehensive analysis would need to take intrinsic alignments into account. However in \cite{Joachimi:2010p855} it was shown that the best calibration for intrinsic alignments comes from the inclusion of galaxy--galaxy lensing with the shear information. Whilst the galaxy--galaxy lensing signal contains a contribution from flux magnification  (Equation \ref{eqn:NumberDensityContrastPS-GGLensing}), we know that the sensitivity of the magnification--ellipticity contribution ($P_{\kappa_{\rm s}m}$) is sub--dominant to the intrinsic clustering--ellipticity correlations ($P_{\kappa_{\rm S} g}$) on all scales and for all redshift bin combinations. Therefore, while the inclusion of a flexible intrinsic alignment model will weaken the constraints for a shear only analysis, we expect that the flux magnification effect will only weakly affect the self--calibration when combining shear and clustering.

\subsection{Tomography}

\begin{figure}
\centering
\includegraphics[width = 0.5\textwidth, trim  = 13mm 15mm 2mm 0mm]{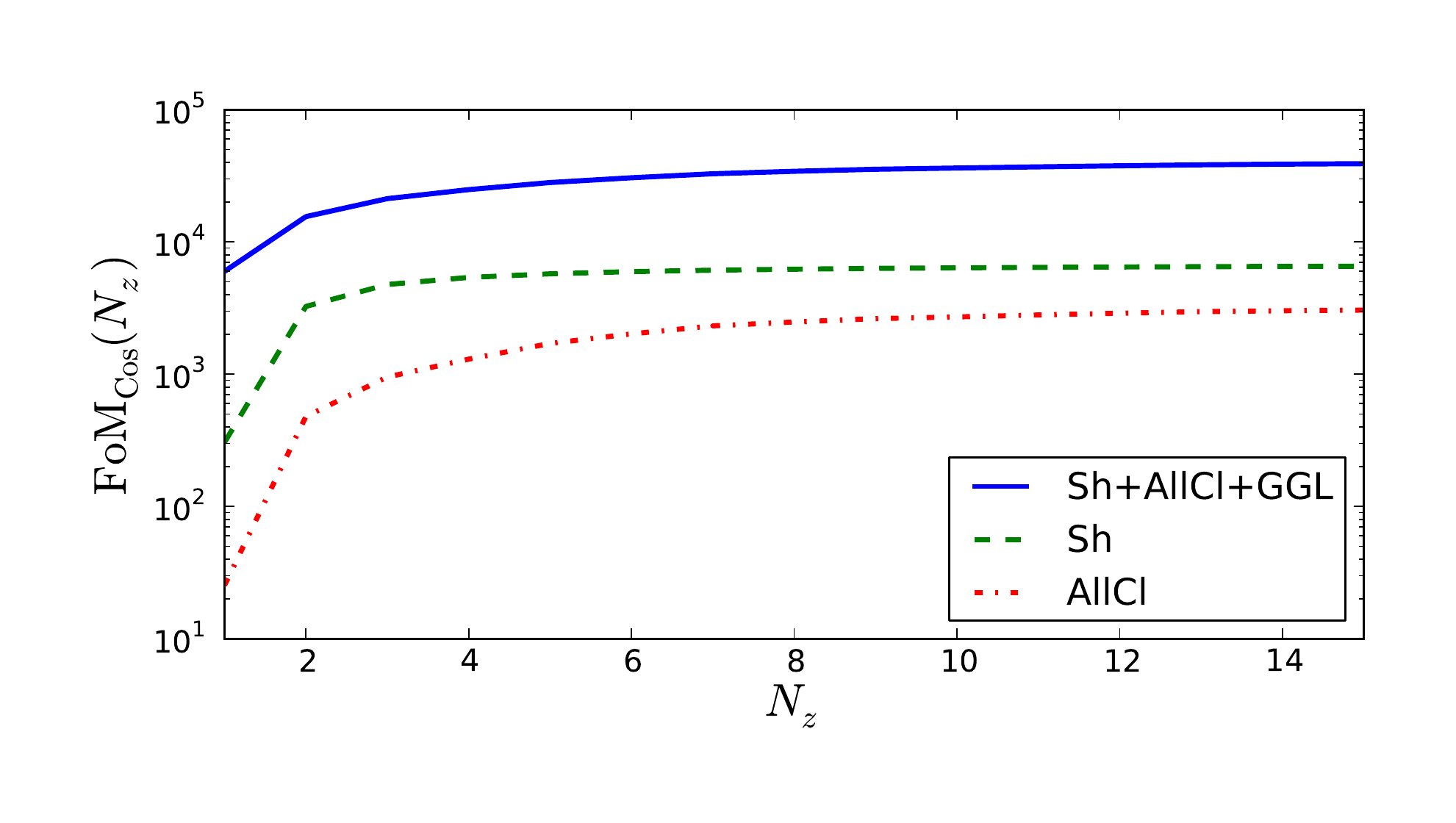}
\caption{Figure of merit (FoM$_{\rm Cos}$) as a function of number of redshift bins, for a S4 survey and for a shear--only (green, dashed), clustering--only (red dot--dashed) and combined (blue, solid) analysis.}\label{Fig:FoM_Nz}
\end{figure}

Here, we consider how much information can be gained from increasing the number of redshift bins we are considering in the analysis. Fig. \ref{Fig:FoM_Nz} shows FoM$_{\rm Cos}$ as a function of $N_z$, the number of redshift bins, for shear--only (Sh), clustering--only (AllCl) and a combined shear, clustering and galaxy--galaxy lensing (Sh+AllCl+GGL) analysis.  In all cases the clustering analysis uses all redshift bin correlations. For all $N_z$, the number of galaxies in each redshift bin is kept the same. We see that for all three cases, the information gain by adding redshift bins is large for small $N_z$, but quickly asymptotes to a constant. We note that for the shear--only case, there is little information gain in taking more than $\sim 4$ redshift bins, which agrees with results shown in \cite{Joachimi:2010p855}. However, the clustering analysis continues to improve constraints up to much higher numbers of redshift bins, and only starts to asymptote to its maximum around $N_z \sim 8$. 
This may be expected as the kernel of the fluctuations for number density due to intrinsic clustering, as defined in Equation \ref{eqn:Projected_NumberDensity}, is much narrower than the lensing kernel, with less overlap between redshift bins, thus we may expect that we can continue to subdivide the galaxy distribution further for the clustering analysis before the clustering power spectra across different bin combinations become highly correlated.

For the combined case, we see that the improvement for a small number of tomographic bins is less than either the shear--only or clustering--only cases. However, constraints from the combined analysis continue to improve up to high $N_z$. By this point, the improvement from shear has asymptoted to its maximum, and the gain comes from continued information gain from the clustering and galaxy--galaxy lensing signals, as the galaxy distribution is further subdivided.

\subsection{Forecasts for various combinations of clustering, cosmic shear and galaxy--galaxy lensing}\label{Sec:AutoClustering_v_Full}

Whilst results presented so far have focused on a clustering analysis which uses all available redshift bin correlations, we may be motivated to consider a clustering analysis which only uses auto--correlations in redshift, as in the auto--correlation terms the intrinsic clustering signal is dominant to the largest magnification contribution, and for which it may be suggested that flux magnification can perhaps be ignored. In this section, we investigate what gains can be made to a photometric redshift clustering analysis when these cross correlations are also included, and whether there is a significant gain when combined with a cosmic shear analysis. Further, we investigate how well galaxy--galaxy lensing can probe cosmology, both as an independent probe or in combination with clustering and cosmic shear.  In this section, we present forecast constraints on dark energy parameters for all analyses set out in Table \ref{Table:Analysis_Types}, and using the Fisher matrix formalism of Section \ref{Sec:ParameterForecasts}, we present forecasts for a S4 survey.

\begin{figure}
\centering
\includegraphics[width = 0.48\textwidth, trim = 0mm 5mm 0mm 0mm]{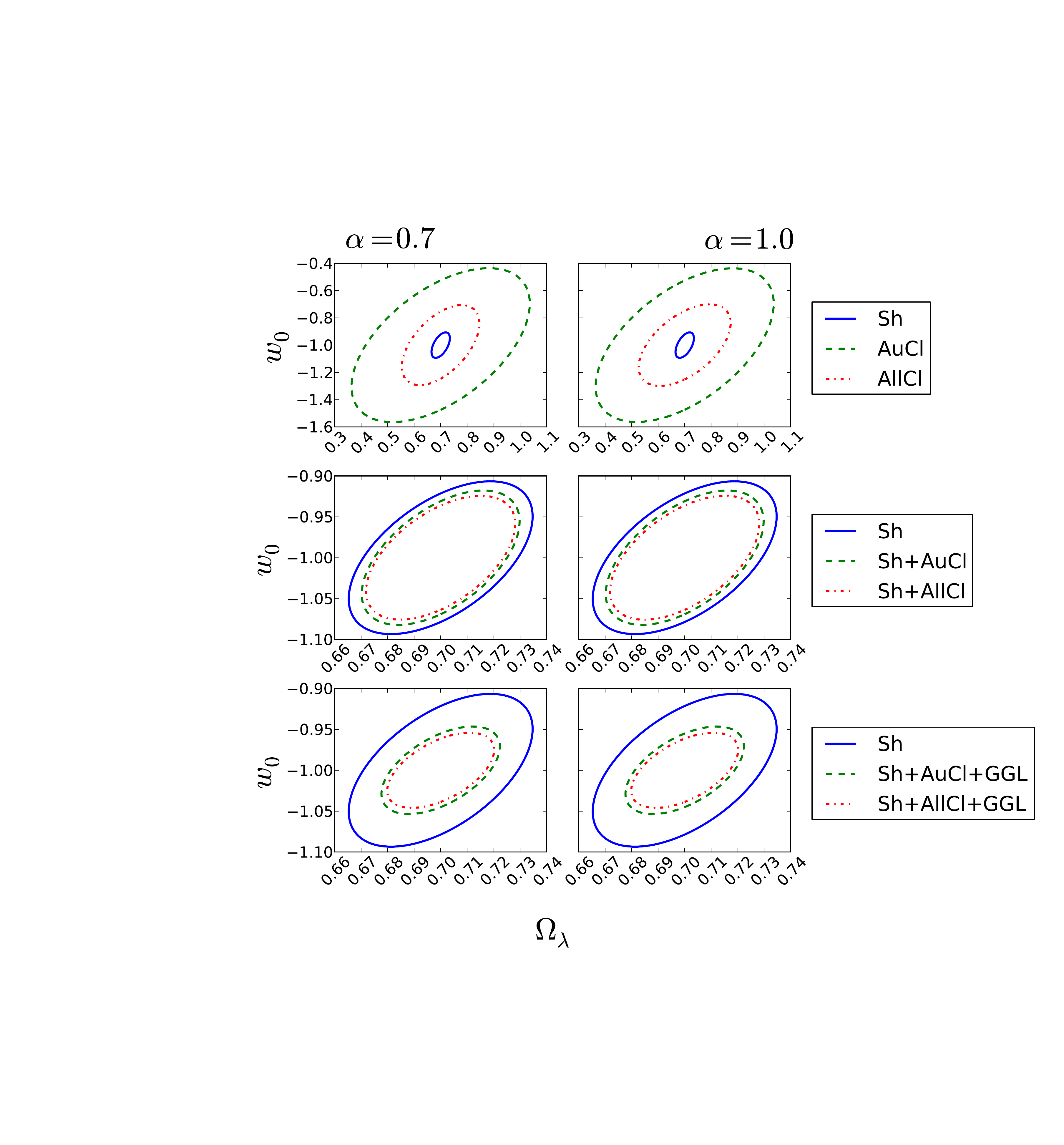}
\caption{Fisher Matrix forecast showing marginal constraints in  $w$, $\Omega_\Lambda$  for an S4 survey, comparing measurements using only the auto--correlation power for the clustering signal (green, dashed), to measurements using all redshift bin combinations in the clustering analysis (red, dot--dashed). All plots show shear--only constraints (blue, solid) for reference, and are the same in all panels. The right column shows contours when $\alpha = 1$, the case when there are no measured correlations in the number density contrast due to flux magnification, and only correlations due to intrinsic clustering. The left column shows contours when $\alpha = 0.7$, as inferred from CFHTLenS catalogues in Section \ref{Sec:CFHT}. Top panels show contours using clustering only, middle panels show clustering combined with cosmic shear, and bottom panels show contours when clustering and galaxy--galaxy lensing information is added to a cosmic shear analysis. All panels take unknown galaxy bias, and cut information from clustering on non--linear scales.}\label{Fig:AutoClustering_v_Full}
\end{figure}

Fig. \ref{Fig:AutoClustering_v_Full} shows marginal constraints in the dark energy parameter ($\Omega_\Lambda, w$) plane, for two different values for $\alpha$. The right column shows contours when $\alpha = 1$, the case when the clustering power spectrum contains no flux magnification contribution. In this case, any improvement in cosmological parameter constraints for a clustering analysis which includes redshift cross--correlations comes from the addition of the intrinsic clustering signal between separated redshift bins, which may be non--zero due to photometric redshift scatter. As well as probing cosmology, these cross--correlations of the intrinsic clustering signal can help constrain galaxy bias, as they are sensitive to cross--correlations between the galaxy bias in different redshift bins. However information from each cross--correlation contribution is expected to be less than the auto term, as in the absence of magnification the power from increasingly widely separated redshift bins rapidly decreases. Similarly, when $\alpha = 1$ the galaxy--galaxy lensing signal reduces to correlations between shear convergence and intrinsic clustering, and the magnification bias contribution vanishes.  In the left column, we present contours for $\alpha = 0.7$, the value inferred from CFHTLenS data as described in Section \ref{Sec:CFHT}. In this case, there is a contribution to the clustering power spectrum from the inclusion of flux magnification effects. Top panels show contours for both types clustering--only analysis (AutoCl, AllCl), middle panels the constraints coming from the addition of cosmic shear information to each clustering analysis (Sh+AutoCl, Sh+AllCl), and bottom panels with the further addition of galaxy--galaxy lensing (Sh+AutoCl+GGL,Sh+AllCl+GGL). In all panels, the cosmic shear contours are shown for reference.

From Fig. \ref{Fig:AutoClustering_v_Full} we see that the use of cross--correlations in redshift bin gives considerable improvements in constraints from a clustering--only analysis. The addition of clustering information to cosmic shear (Sh+AutoCl, Sh+AllCl) improves constraints over the shear--only case, with further significant improvement with the addition of galaxy--galaxy lensing. 

These results are borne out by Tables \ref{Table:FoM_DE_byAnalysis_S4} and \ref{Table:FoM_DE_byAnalysis_S3}, which show FoM$_{\rm DE}$ values, with $1\sigma$ errors on $\Omega_\Lambda$ and $w$ for an S4  and S3 survey respectively. We see a significant improvement in parameter constraints with the addition of redshift cross--correlations in a clustering--only analysis (AutoCl to AllCl) for both survey types. For a S4 survey which takes galaxy bias as unknown and $\alpha = 0.7$, the use of clustering cross--correlations shows a decrease by a factor of $\sim 2$ in the statistical error of $w$, $\sim 2.26$ in $\Omega_\Lambda$, and an increase in FoM$_{\rm DE}$ by a factor of $3.2$ over the AutoCl case.  When $\alpha = 1$, FoM$_{\rm DE}$ shows an increase by a factor of $2.4$.

The combination of both clustering analyses with galaxy--galaxy lensing (AutoCl+GGL, AllCl+GGL) both outperform a cosmic shear only analysis (Sh), confirming that the combination of clustering with galaxy--galaxy lensing provides a competitive alternative to cosmic shear, as suggested by \cite{Mandelbaum21062013}.

The combination of clustering redshift auto--correlations with cosmic shear (Sh+AutoCl) show an improvement on parameter errors of $\sim 15\%$ for both dark energy parameters over the shear--only (Sh) case, corresponding to a $33\%$ increase in FoM$_{\rm DE}$. The addition of redshift cross--correlations in the clustering analysis (Sh+AllCl) gives a further improvement of $14\%$ in FoM$_{\rm DE}$.

For an analysis which combines cosmic shear, galaxy--galaxy lensing and clustering using all redshift bin correlations (Sh+AllCL+GGL), there is a significant improvement over parameter constraints when compared to the cosmic--shear--only case (Sh), with a decrease of $\sim 2.0$ and $\sim 1.75$ in statistical errors on $w$ and $\Omega_\Lambda$ respectively, giving an improvement by a factor of $3.7$ to FoM$_{\rm DE}$. This is a $\sim 30\%$ improvement to the case where only clustering auto--correlations in redshift are combined with cosmic shear and galaxy--galaxy lensing (Sh+AutoCl+GGL).
 
We note that there is little difference in statistical errors between both choices of $\alpha$ for the various combinations of shear with clustering (Sh+AllCl, Sh+AutoCl), and even if galaxy--galaxy lensing is also included (Sh+AllCl+GGL, Sh+AutoCl+GGL). For shear with clustering auto--correlations (Sh+AutoCl), there is no change to FoM$_{\rm DE}$ between both choices of $\alpha$, as the clustering auto--correlation terms are dominated by the intrinsic clustering signal. With the inclusion of clustering cross--correlations in redshift (Sh+AllCl), the presence of flux magnification gives a sub--percent change to the dark energy Figure of Merit, increasing to percent level for the case of shear combined with all clustering information and galaxy--galaxy lensing (Sh+AllCl+GGL). This agrees with the results detailed in Section \ref{Sec:Constraints_AlphaDependance}. 

We therefore conclude that there is significant gain in considering all redshift bin combinations in the clustering signal rather than the redshift auto--correlations only. 
The addition of clustering to cosmic shear gives noticeable improvement in parameter constraints, with further significant improvement with the addition of galaxy--galaxy lensing. In each case, the use of all redshift bin correlations in the clustering analysis gives modest improvement to constraints. We find that the flux magnification signal significantly reduces the error in clustering--only constraints, but it reduces statistical errors in a joint shear--clustering analysis only minimally.
\subsection{Biases in cosmological parameter estimates}\label{Sec:InferredParameterBiases}

\begin{table*}
\begin{center}
\caption{Table showing how constraints on Dark Energy parameters (through FoM$_{\rm DE}$) vary as galaxy bias is perfectly known (fixed) or constrained with the data (free) for an S4 survey. Columns labelled with $\alpha$ values in the header show FoM$_{\rm DE}$ values. The last two columns show $1\sigma$ constraint for $\Omega_{\Lambda}$ and $w$ when galaxy bias is free and taking $\alpha = 0.7$. Columns labelled $\Delta_{\Omega_\Lambda}$ and $\Delta_w$ show bias on dark energy parameters, when flux magnification is neglected (when $\alpha = 1$ is incorrectly assumed). For each analysis considered in Table \ref{Table:Analysis_Types}, results are presented for the case with no flux magnification ($\alpha = 1$) and using $\alpha = 0.7$ as measured from CFHTLenS catalogues (Section \ref{Sec:CFHT}). }\label{Table:FoM_DE_byAnalysis_S4}
\begin{tabular}[]{l | c | c | c | c |c|c|c|c|}
\hline
S4& \multicolumn{2}{|l|}{$b$ fixed} & \multicolumn{6}{|l|}{$b$ free} \\
\hline
											& $\alpha = 1$ & $\alpha = 0.7$ & $\alpha = 1$ & $\alpha = 0.7$& $\sigma_{\Omega_\Lambda}$& $\Delta_{\Omega_\Lambda}$ & $\sigma_{w}$ & $\Delta_{w}$\\
\hline
AutoCl				  & 							$582$& 			$621$ & 		$12.4$ & 		$12.4$	& $0.34$ 		&0.026		& $0.56$		& 0.045 \\
AllCl 			 & 							$712$& 			$1965$ & 		$30.1$ & 		$39.5$	& $0.15$ 		&0.69		& $0.29$		&0.77 \\
GGL &										161&			260&		49.0&		84.0		& 0.11		&-0.076		& 0.11		&-0.044 \\
AutoCl + GGL &									5891&			5976&		719&		736		& 0.033		&-0.12		& 0.071		&-0.50 \\
AllCl + GGL &									6964&			7110&		919&		946		& 0.030		&-0.2		& 0.062		&-0.58 \\
Sh		  & 									$685$& 			$685$ & 		$685$ & 		$685$	& $0.035$ 	&-			& $0.093$		&- \\
Sh + GGL  &									2134&			2155&		1324&		1335		& 0.029		&-0.12		& 0.067		&-0.39 \\
Sh + AutoCl				 & 						$4896$& 			$5007$ & 		$914$ & 		$914$	& $0.030$		& 0.005		& $0.082$		& $0.013$	\\
Sh + AllCl 		 & 							$5514$& 			$5634$ & 		$1032$ & 		$1040$	& $0.028$ 	&-0.03		& $0.076$		& $0.12$	\\
Sh + AutoCl + GGL &								7046&			7139&		1927&		1935		& 0.022		&0.19		&0.054		&0.24 \\
Sh + AllCl + GGL 						& 		$8264$& 			$8426$ & 		$2468$ & 		$2505$	& $0.020$		& 0.11		& $0.046$		& $0.13$	\\
\hline
\end{tabular}
\end{center}
\end{table*}

\begin{table*}
\centering
\caption{As Table \ref{Table:FoM_DE_byAnalysis_S4}, for an S3 survey.}\label{Table:FoM_DE_byAnalysis_S3}
\begin{tabular}[]{l | c | c | c | c |c|c|c|c|}
\hline
S3& \multicolumn{2}{|l|}{$b$ fixed} & \multicolumn{6}{|l|}{$b$ free} \\
\hline
									& $\alpha = 1$ & $\alpha = 0.7$ & $\alpha = 1$ & $\alpha = 0.7$& $\sigma_{\Omega_\Lambda}$ & $\Delta_{\Omega_\Lambda}$& $\sigma_{w}$ & $\Delta_{w}$\\
\hline
AutoCl  & 								$44.5$& 			$47$ & 		$0.87$ & 		$0.86$	& $1.32$ 		&0.02			& $2.16$ 		& 0.033 \\
AllCl  & 								$54.0$& 			$144$ & 		$2.2$ & 		$2.8$	& $0.57$ 		&0.72 			& $1.10$ 		& 0.77 \\
GGL &								5.5&				8.9&			1.47&		2.3		& 0.66		&-0.028			& 0.65		&-0.1 \\
AutoCl + GGL &							330&			338&		32.0&		32.2		& 0.17		&0.1				& 0.4			&-0.086 \\
AllCl + GGL &							375&			386&		39.0&		39.6		& 0.15		&-0.024			& 0.35		&-0.24 \\
Sh  & 								$10.4$& 			$10.4$ & 		$10.4$ & 		$10.4$	& $0.3$ 		& -				& $0.8$ 		& - \\
Sh + GGL  &							42.6&			43.3&		24.4&		24.7		& 0.21		&-0.17			& 0.46		&-0.52 \\
Sh+ AutoCl & 							$273$& 			$281$ & 		$18.3$ & 		$18.3$	& $0.22$		& 0.0066			& $0.56$		& 0.022 	\\
Sh + AllCl  & 							$314$& 			$322$ & 		$21.5$ & 		$21.8$	& $0.20$		&-0.038			& $0.50$ 		& 0.32	\\
Sh + AutoCl + GGL &						349&			357&		52.8&		53.1		& 0.14		&0.28			&0.32		&0.36 \\
Sh + AllCl + GGL & 						$399$& 			$410$ & 		$64.1$ & 		$65.3$	& $0.13$		& 0.14			& $0.28$		& $ 0.20$	\\
\hline
\end{tabular}
\end{table*}

In Section \ref{Sec:AutoClustering_v_Full} we found that the {\it precision} of results for the various combinations of clustering information with cosmic shear and galaxy--galaxy lensing is not significantly improved by the presence of a magnification signal. In this section, we investigate how this flux magnification signal affects the {\it accuracy} of each analysis. We therefore turn our attention away from constraints on cosmological parameters, and instead consider possible shifts in the deduced maximum likelihood point when data are fitted using only the intrinsic clustering power spectrum, and where flux magnification has been neglected. To do this, we consider the linear shift in inferred parameters $(Q)$ due to a bias in fixed model parameters ($\psi$) using the formalism of \cite{Taylor:2007p1037}
\be\label{eqn:ParameterBias}
\bm{\delta Q}_i = -\sum_{k,j}[\mathbfss{F}^{QQ}]^{-1}_{ik}\mathbfss{F}^{Q\psi}_{kj}\bm{\delta \psi}_j,
\ee
where $\mathbfss{F}^{QQ}$ is the Fisher matrix of measured  cosmological parameters as set out in Section \ref{Sec:ParameterForecasts},  and $\mathbfss{F}^{Q\psi}$ is a pseudo--Fisher matrix between inferred and fixed model parameters using the same formalism. Our measured parameters consist of the cosmological parameter set described in Section \ref{Sec:ParameterForecasts} including $N_z$ galaxy bias parameters (where $N_z$ is the number of redshift bins). Our set of assumed parameters consists of one $\alpha$ value per redshift bin, so that $\delta \psi^i = \alpha^i_{\rm true} -  1$, where we have noted that  setting $\alpha = 1$ sets the contribution to number density contrast fluctuations from magnification identically zero, thus equivalent to fitting only using the intrinsic clustering power spectrum. As previously, we choose $\alpha_{\rm true} = \ATrue$. The same formalism calculates the bias in parameter estimates arising from an incorrect estimation of $\alpha$.

The columns of Tables  \ref{Table:FoM_DE_byAnalysis_S4} and \ref{Table:FoM_DE_byAnalysis_S3} labelled with $\Delta_{\Omega_\Lambda}$ and $\Delta_{w}$ show the bias on each dark energy parameter introduced by incorrectly fitting to the clustering analysis ignoring the flux magnification contribution to the clustering power spectrum. We note that biases in cosmological parameters are reasonably robust to the survey type. However as the S4 survey gives better constraints to parameter values, biases in cosmological parameters are more significant for the S4 survey type than for the S3 case. 

For a clustering analysis using only auto correlations in redshift, we find that biases in all cosmological parameters from incorrectly ignoring flux magnification are smaller than statistical errors. The inclusion of cross--correlations in a clustering analysis increases the size of bias on parameters, as the flux magnification contribution to the power spectrum increases with more widely separated redshift bins (see Fig. \ref{Fig:NumberPowerSpectra}). As the induced parameter biases can be at least as large as the statistical errors, particularly for the S4 survey, it is clear that whilst there is a significant increase in the figure of merit for a clustering analysis by including cross--correlations in redshift, this comes at the expense of increased complexity as flux magnification must be accurately measured and modelled to avoid large biases in inferred cosmological parameters. 

Using the bias on $\Omega_{\rm M}$, chosen as the most significant case for AllCl and having a value of $\Delta_{\Omega_{\rm M}} = -0.39$, we find that $\alpha$ must be measured to within $\Delta \alpha  = 0.085$ to ensure biases are smaller than statistical errors for all parameters for an S3 survey. For a S4 survey, this becomes $\Delta \alpha  = 0.025$. The value of $\alpha$ measured from CFHTLenS data is known to within percent level precision at the faint limit of the survey. However this statistical uncertainty does not factor in expected systematic shifts in $\alpha$ due to selection bias, completeness or other observational effects. 

Provided these effects are small, biases in inferred parameters should be smaller than statistical errors. However if these effects cause percent level shifts in the measured value of $\alpha$, a photometric clustering analysis including redshift bin cross--correlations may be significantly biased even if the flux magnification effect is measured and modelled. This could be partially mitigated if statistical errors on $\alpha$ are propagated into cosmological parameter constraints, causing a decrease in figure of merit.

A combined shear and clustering analysis (Sh+AllCl, Sh+AutoCl) shows smaller biases than its clustering--only counterparts as the use of cosmic shear information, which is unaffected by flux magnification, helps to constrain parameter values. However, the further addition of galaxy--galaxy lensing, whose power spectrum includes a flux magnification contribution, causes a further increase to parameter biases to values larger than statistical errors for both the S3 and S4 survey types. We note that the bias on inferred parameters for the combination of clustering auto--correlations with cosmic shear and galaxy--galaxy lensing is of the same order of magnitude as the bias when combined with clustering using all redshift bin combinations (Sh+AllCl+GGL). Therefore, whilst there is an important improvement to the figure of merit when clustering and galaxy--galaxy lensing is added to shear (Sh+AutoCl+GGL, Sh+AllCl+GGL), as with the clustering--only case this comes at the expense of increased complexity as flux magnification must be accurately measured and modelled to avoid biasing inferred parameters for both analyses. This suggests that all available information should be used to maximise the strength of parameter constraints, and Sh+AllCl+GGL should be used in preference to Sh+AutoCl+GGL as $\alpha$ would need to be accurately measured in either case.  For the Sh+AllCl+GGL case, which shows the strongest constraining power, the bias on $\Omega_\Lambda$ is the most significant of all cosmological parameters, and using this value we find that $\alpha$ must be measured to within $\Delta\alpha = 0.13$ for an S3 survey, and $\Delta\alpha = 0.055$ for an S4 survey to keep biases smaller than statistical errors on each parameter.

\subsection{The effect of cuts on non--linear scales}\label{Sec:ell-cut-dependance}

As detailed in Section \ref{Sec:GalaxyBiasModelling}, we impose cuts on $\ell$--modes of the data vector where a significant part of the clustering signal is expected to come from regions where the galaxy bias is non--linear, thus allowing us to restrict our attention only the the linear regime where galaxy bias can be modelled well. In the main body of this text, we have assumed that cuts following the method laid out in Section \ref{Sec:GalaxyBiasModelling} with $\sigma_R = 0.5$ are adequate, as this gives a maximum $k$--mode within the quasi--linear regime (see for example \citealt{Cole2dFRGS}, which shows only a few percent deviation from linear theory on these scales). In this Section, we investigate how the results presented change with a change in $\sigma_R$, with particular emphasis on a more conservative cut using $\sigma_R = 0.2$, which gives $k_{\rm max} = 0.1\;h/{\rm Mpc}$ at z = 0.

\begin{figure}
\centering
\includegraphics[width = 0.48\textwidth, trim = 0mm 5mm 0mm 0mm]{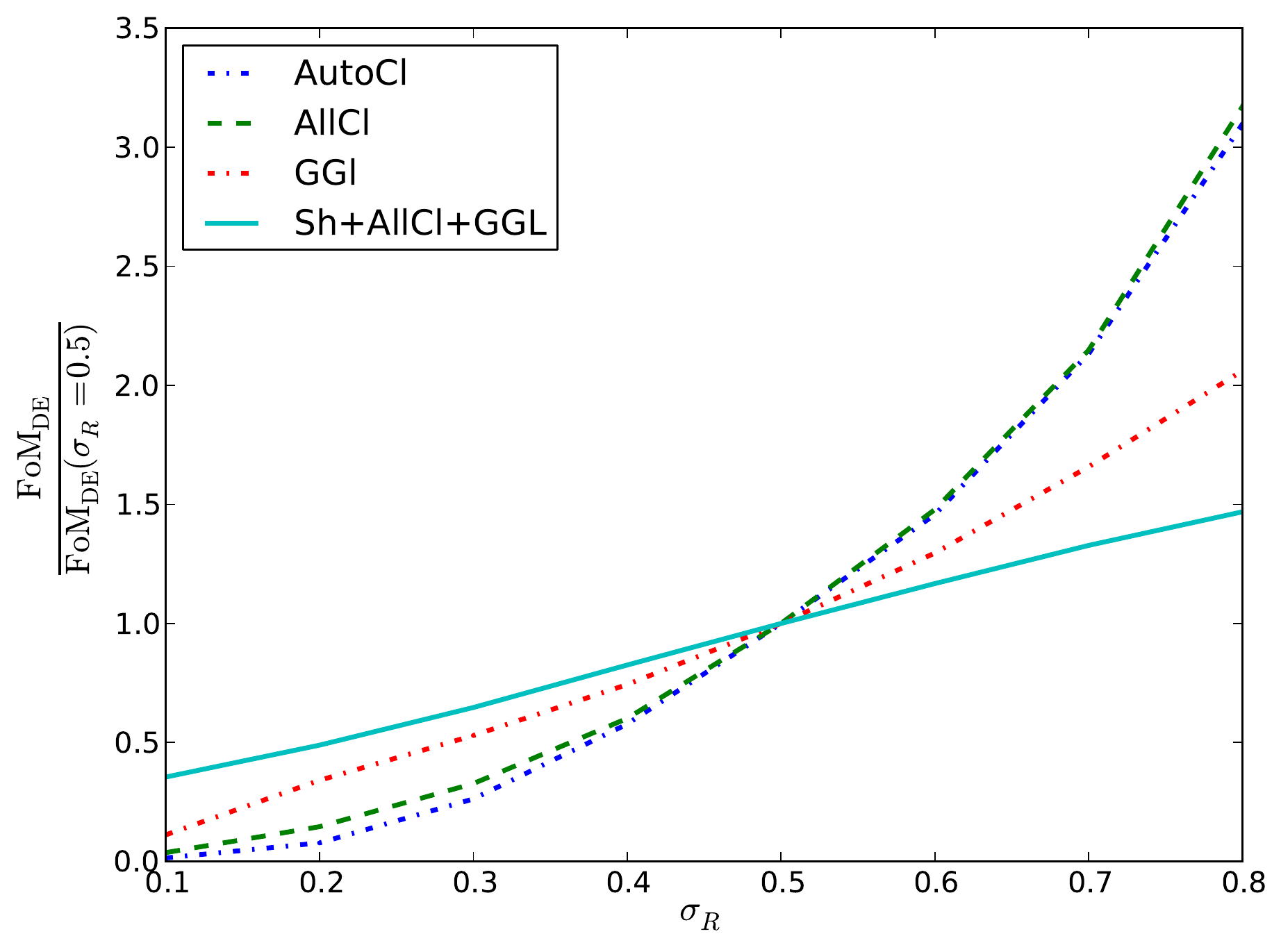}
\caption{Figure showing how the dark energy figure of merit FoM$_{\rm DE}$ varies as a function of $\sigma_R$, which sets maximum $\ell$--mode to be included in a clustering or galaxy--galaxy lensing analysis. The method used to apply these cuts are defined in Section \ref{Sec:GalaxyBiasModelling}. The ordinate axis has been renormalised to the figure of merit value when $\sigma_R = 0.5$, the value used in the main body of this text, for each analysis considered.}\label{Fig:FoM_vs_sigR}
\end{figure}

Fig. \ref{Fig:FoM_vs_sigR} shows how the dark energy figure of merit, FoM$_{\rm DE}$, renormalised to its value when $\sigma_R = 0.5$, varies for a selection of analyses that include number density information, as a function of $\sigma_R$. Both clustering analyses, AutoCl and AllCl, depend strongly on the scale set for non--linear cuts, with the figure of merit an order of magnitude smaller when the more conservative $\sigma_R = 0.2$ is used. It is notable that the behaviour of the FoM for both the AutoCl and AllCl cases are very similar, suggesting that most of the improvement in FoM due to the inclusion of more non--linear scales comes from the auto--correlations in number density, and particularly the intrinsic clustering contribution. The information from a galaxy--galaxy lensing (GGL) analysis depends less strongly on the scale of non--linear cuts than either of the clustering--only analyses. Finally, the full combined probe, Sh+AllCl+GGL, shows the weakest dependence on cuts on non--linear scales, as no cuts are applied to the shear information. We note however that in all cases, more conservative cuts give a noticeable change in the cosmological constraints for each analysis, as more conservative cuts discard more information from which cosmology can be inferred. For the full combined probe, choosing $\sigma_R = 0.2$ gives a reduction by a factor of $\approx 2$ in the figure of merit compared to the less conservative choice of $\sigma_R = 0.5$. 

However, qualitatively the results presented in this text do not change with the choice of either $\sigma_R = 0.2$ or $0.5$. In particular, with the more conservative $\ell$--cuts the magnification contribution to a full combined analysis gives only sub--percent change to the figure of merit for realistic choices of $\alpha$, in agreement with the discussion in Section \ref{Sec:Constraints_AlphaDependance}. Further, conclusions on biases on inferred parameters (Section \ref{Sec:InferredParameterBiases}) remain unchanged with either choice, however it should be noted that biases on inferred parameters caused by neglecting the magnification signal tend to be smaller with more conservative cuts due to the reduced constraining power of the clustering and galaxy--galaxy lensing signal. Further, these biases  are usually less significant due to the increased statistical error on each parameter when using more conservative $\ell$--cuts.

\section{Conclusions}\label{Sec:Conclusions}

In this paper we have made forecasts for statistical errors on cosmological parameters from a cosmic shear analysis, and two types of  clustering analysis using photometric redshift information, including flux magnification effects: one which contains information only from correlating all redshift bins with themselves (auto); and one where all redshift bin combinations are included, both in the presence of scatter due to photometric redshift errors. We considered the gain from combining either clustering analysis with a cosmic shear and/or galaxy--galaxy lensing, and investigated how inferred cosmological parameters are biased when the flux magnification effect is neglected. The full list of experiments considered is given in Table \ref{Table:Analysis_Types}.

Using public CFHTLenS catalogues, we inferred a value for the slope of the cumulative number counts, $\alpha$, at the faint limit of the survey to be $\alpha \approx 0.7$. Although cuts at brighter magnitude limits can give a larger value of $\alpha$, we find that the subsequent removal of sources from our sample causes a decrease in signal--to--noise ratio in all but the highest redshift bins where photometry is least accurate. 

For a Stage--IV type survey, the inclusion of all redshift correlations in a clustering--only analysis can give significant improvement to cosmological constraints, improving FoM$_{\rm DE}$ by a factor of $3.2$ over a clustering analysis which uses only auto--correlations in redshift. When combined with a shear analysis, clustering using only the auto--correlations in tomographic redshift bins improves the figure of merit for dark energy parameters $\Omega_\Lambda$,$w_0$ (FoM$_{\rm DE}$) by a factor of $1.33$ over the shear--only case. Including cross--correlations improves FoM$_{\rm DE}$ by a factor of $1.51$ over a shear--only analysis. However if galaxy--galaxy lensing is also included this increases to $3.7$ when galaxy bias is unknown. In contrast, when galaxy bias is perfectly known, the addition of clustering information using all redshift bin correlations and galaxy--galaxy lensing improves FoM$_{\rm DE}$ by a factor of $12.3$ over a shear--only analysis.

Whilst a non--zero flux magnification can cause a significant improvement in constraints from a clustering--only analysis when galaxy bias is free (giving an increase in figure of merit by a factor of $\sim 1.3$ for clustering using all redshift bin combinations, when $\alpha = 0.7$ compared to $\alpha = 1$), the presence of a magnification signal in the clustering data causes only a few percentage change in parameter constraints from a joint clustering, shear and galaxy--galaxy lensing analysis and does not significantly alter the precision of results. However, if clustering cross--correlations in redshift or galaxy--galaxy lensing are used as part of an analysis, the flux magnification effect must be modelled and included to avoid inferred parameter constraints being largely biased and inaccurate. 

If flux magnification is incorrectly neglected, biases in inferred cosmological parameters from clustering--only analysis which includes cross--correlations in redshift can be many times larger than statistical uncertainties for an S4 survey. For a combined clustering, cosmic shear and galaxy--galaxy lensing analysis, parameter biases can be larger than statistical uncertainties for an S4 survey, and that there is little improvement in these biases when clustering redshift cross--correlations are discarded. As there is significant improvement in the statistical errors when clustering redshift cross--correlations are used in this type of analysis, we suggest that these are included. This comprises the case where all available information is used from number density contrast and ellipticity measurements.

As the use of all available information in this way outperforms all other analyses (see Table \ref{Table:FoM_DE_byAnalysis_S4}) we suggest that given number density information and ellipticity measurements, the full combined analysis which includes cosmic shear, galaxy--galaxy lensing and clustering using all redshift bin combinations should be used (Sh+AllCl+GGL in Table \ref{Table:FoM_DE_byAnalysis_S4}). When compared to a cosmic shear--only analysis, this comes at the expense of increased complexity due to the necessity to accurately measure and model flux magnification to avoid causing these parameters to be biased. Alternatively, the combination of cosmic shear and number density contrast redshift auto--correlations (Sh+AutoCl) shows the smallest bias in inferred parameters when flux magnification is neglected, and thus may reduce the complexity of the analysis at the expense of weaker parameter constraints.

As well as the improvement in statistical errors that comes from adding clustering measurements and galaxy--galaxy lensing to cosmic shear, number density information gives an important consistency check by allowing independent verification of inferred parameter values using either number density contrast correlations, or galaxy--galaxy lensing. As the information required for a photometric clustering analysis is already taken as part of a shear survey, this information is obtained without the need for further data, provided $\alpha$ can be accurately measured.

Code available at https://github.com/ChristopherAJDuncan/ClusteringShearComplementarity.

\section*{Acknowledgments}

The authors would like to thank the anonymous referee, whose comments helped improve the paper. Christopher Duncan is supported by an STFC scholarship. Benjamin Joachimi acknowledges support by an STFC Ernest Rutherford Fellowship, grant reference ST/J004421/1. Alan Heavens thanks the IfA for travel support. Catherine Heymans acknowledges support from the European Research Council under the EC FP7 grant number 240185. Hendrik Hildebrandt is supported by the Marie Curie IOF 252760, a CITA National Fellowship, and the DFG grant Hi1495/2-1. 


We thank the CFHTLenS team for making their catalogues publicly available to download from www.cfhtlens.org.  This work is partly based on observations obtained with MegaPrime/MegaCam, a joint project of CFHT and CEA/IRFU, at the Canada--France--Hawaii Telescope (CFHT) which is operated by the National Research Council (NRC) of Canada, the Institut National des Sciences de l'Univers of the Centre National de la Recherche Scientifique (CNRS) of France, and the University of Hawaii. This research used the facilities of the Canadian Astronomy Data Centre operated by the National Research Council of Canada with the support of the Canadian Space Agency. CFHTLenS data processing was made possible thanks to significant computing support from the NSERC Research Tools and Instruments grant program.

\bibliographystyle{mn2e}
\setlength{\bibhang}{2.0em}
\setlength\labelwidth{0.0em}
\bibliography{Papers_BIBTEX}

\appendix

\section[]{The Galaxy Bias Prior -- Normalising the Likelihood}\label{App:NormalisingGBP}

We define the covariance matrix for the prior on $N_z$ linear galaxy bias parameters as in Equation (\ref{Eqn:BiasPriorCov}). The covariance matrix then takes the form $\mathbfss{C} = \sigma_\nu^2 \mathbfss{R}$, where $\mathbfss{R}$ is the matrix of correlation parameters ($\nu$) in Toeplitz form. Assuming the full likelihood for the bias parameters is Gaussian, it then follows that
\be
\frac{p(\bar{b})}{p(\bar{0})}  =  e^{-\frac{1}{2}\chi^2},
\ee
 where $\bar{b}$ labels the set of galaxy bias nuisance parameters, we have renormalised to the likelihood when $\bar{b} = \bar{0}$ to remove any pre-factors, and where
 \be
 \chi^2  = \bm{b}\mathbfss{C}^{-1}\bm{b}^T =  \sum_{ij} ^{N_z} b^2 (\mathbfss{C}^{-1})_{ij},
 \ee
 along the line $b_1 = b_2 = \cdot\cdot\cdot = b_{N_z} \equiv b$. From the definition of the covariance matrix in Equation (\ref{Eqn:BiasPriorCov}) it follows that
  \begin{eqnarray}
   \chi^2 &=& \frac{b^2}{\sigma_\nu^2}\sum_{ij}^n (\mathbfss{R}^{-1})_{ij} = \frac{b^2}{\sigma_\nu^2}\sum_{ij}^{N_z} \left[\frac{\mathbfss{Adj}(\mathbfss{R})}{{\rm det}(\mathbfss{R})}\right]_{ij}, \nonumber\\ &=& \frac{b^2}{\sigma_\nu^2 {\rm det}(\mathbfss{R})}\sum_{ij}^{N_z} [\mathbfss{Co}(\mathbfss{R})^{\rm T}]_{ij},\nonumber\\
 &=&  \frac{b^2}{\sigma_\nu^2 {\rm det}(\mathbfss{R})}\sum_{ij}^{N_z} [ \mathbfss{Co}(\mathbfss{R})]_{ij}, \label{eqn:nChi}
 \end{eqnarray}
 where $\mathbfss{Adj}(\mathbfss{R})$ denotes the adjoint matrix of $\mathbfss{R}$, $\mathbfss{Co}(\mathbfss{R})$ the matrix of cofactors, defined as $ {\rm Co(\mathbfss{R})}_{ij} = (-1)^{i+j}{\rm M}_{ij}$, $\mathbfss{M}$ the minor matrix of determinants, and we have used the symmetry of $R$ to note that $\mathbfss{Co}(\mathbfss{R})^{\rm T} = \mathbfss{Co}(\mathbfss{R})$. 
  
By the symmetry of $\mathbfss{R}$, and the definition of the minor matrix of determinants, the matrix of cofactors  $\mathbfss{Co}(\mathbfss{R})$ satisfies 
\be
\mathbfss{Co}(\mathbfss{R}) = \begin{pmatrix} x_1 & x_2 & 0 & 0 & \cdots & 0 \\  x_2 & x_3 & x_2 & 0 & \cdots & 0 \\ 0 & x_2 & x_3 & x_2 & \cdots & 0 \\ \vdots &  & \ddots &  &  & \vdots \\  0 & \cdots & 0 & 0 & x_2 & x_1 \end{pmatrix},
\ee
so that
 \be\label{eqn:SumCo}
 \sum_{ij}^n \mathbfss{ Co}(\mathbfss{R})_{ij} = 2x_1 + 2(N_z-1)x_2 + (N_z-2)x_3,
 \ee
 and ${\rm det}(\mathbfss{R}) = (1-\nu^2)^{N_z-1}$.  Variable $x_1$ is the determinant of the $N_z-1$ case of the matrix $\mathbfss{R}$, which we denote as ${\rm det}(\mathbfss{R}^{(N_z-1)})$. The remaining factors can be calculated by noting that
 \be
 \sum_{j}^{N_z} \mathbfss{R}_{ij} \mathbfss{Co}(\mathbfss{R})_{ij} = {\rm det}(\mathbfss{R}),
 \ee
 so that 
 \begin{eqnarray}
x_2 &=& \frac{1}{\nu}[{\rm det}(\mathbfss{R}^{(N_z)}) - {\rm det}(\mathbfss{R}^{(N_z-1)})], \\
x_3 &=& 2{\rm det}(\mathbfss{R}^{(N_z-1)}) - {\rm det}(\mathbfss{R}^{(N_z)}).
\end{eqnarray}

Using Equations (\ref{eqn:SumCo}) and (\ref{eqn:nChi}), we find that 
\be
 \chi^2 = \frac{b^2}{\sigma_\nu^2}\left[\frac{N_z-(N_z-2)\nu}{1+\nu}\right].
 \ee
It is clear then that if we keep $\sigma_\nu$ constant as we change $\nu$, $\chi^2$ at a given point along the $b_1 = b_2 = \cdot\cdot\cdot = b_{N_z} = b$ line will vary with $\nu$ causing a change in the volume of galaxy bias parameter space probed by $1\sigma$ contours to also vary with the correlation strength. We therefore normalise $\chi^2$ along $b_i = b$ by choosing $\sigma_\nu$ such that $\chi^2_\nu = \chi^2_{\nu=0}$ for all values of $\nu$. Noting that by Pythagoras' theorem $b = \sigma_0/\sqrt{N_z}$ and requiring that $\sigma_\nu(\nu=0) = \sigma_0$, we find that
 \be
 \sigma_\nu = \sigma_0\left[\frac{N_z-(N_z-2)\nu}{N_z(1+\nu)}\right]^{\frac{1}{2}},
 \ee
  as stated in Equation (\ref{eqn:BP_sigr_sig0}).

\section{An alternative Fisher matrix analysis for Sh, AllCl, and Sh+AllCl+GGL}\label{Sec:FM_1pt}
For the cases where we wish to use the full information for a given probe, such as using ellipticity information, number over-density or a combination of both over all redshift bins, we can alternatively take the data vector to contain either ellipticity, number density contrast, or both.  By definition, the data then have zero mean, and the Fisher matrix elements for parameters labelled by $\eta$ and $\tau$, takes the form \cite{Tegmark:1997p9} as
\be
\mathbfss{F}_{\eta\tau} = \sum_{\bm{\ell}_r}\frac{\bm{\ell}_r\Delta\bm{\ell}_r\Delta\Omega}{4\pi}Tr[\mathbfss{C}^{-1}(\bm{\ell}_r)\mathbfss{C}_{,\eta}(\bm{\ell}_r)\mathbfss{C}^{-1}(\bm{\ell}_r)\mathbfss{C}_{,\tau} (\bm{\ell}_r)],
\ee 
 where $\mathbfss{C}$ is a matrix containing the covariance between all  $1$-point estimators entering the data vector. 

Using this formalism, the covariance matrix takes a simpler form than for the formalism detailed in Section \ref{Sec:ParameterForecasts} . For example, for the combination of a shear analysis with information from clustering and magnification over all redshift bins, and including galaxy-galaxy lensing (Sh+AllCl+GGL), the data vector takes the form $\bm{D}(\bm{\ell}) = \{\epsilon^{(1)}(\bm{\ell}),\ldots,\epsilon^{(N_z)}(\bm{\ell}),\delta n^{(1)}(\bm{\ell}),\ldots,\delta n^{(N_z)}(\bm{\ell})\}$ so that the covariance matrix takes block form
\be
\mathbfss{C}(\bm{\ell}) = \left(\begin{array}{c|c} P_{\epsilon\epsilon}(\bm{\ell})&P_{\epsilon \delta n}(\bm{\ell})\\\hline P_{\delta n \epsilon}(\bm{\ell})&P_{\delta n\delta n}(\bm{\ell})\end{array}\right),
\ee
with power spectra as defined in Equations (\ref{eqn:NumberDensityContrastPS} through \ref{eqn:NumberDensityContrastPS-GGLensing}). 

The equivalence of both the above prescription and the prescription set out in Section \ref{Sec:ParameterForecasts} is detailed in \cite{Joachimi:2011p2014} when assuming Gaussian statistics. Results for the Sh, AllCl and Sh+AllCl+GGL analyses were verified using both methods for this paper.

\bsp

\label{lastpage}

\end{document}